\documentclass[]{aa}
\usepackage[varg]{txfonts}
\usepackage{graphicx}
\usepackage{booktabs}
\usepackage{amsmath}
\usepackage{colortbl}
\usepackage{subfig}
\usepackage{rotating}
\usepackage{amssymb}
\usepackage{latexsym}
\usepackage{ifthen}
\usepackage{enumitem}

\begin{document}

\title{The molecular gas content in obscured AGN at z$> 1$} 

\author{M. Perna
		\inst{\ref{i1},\ref{i2}}\thanks{E-mail: perna@arcetri.inaf.it}
		\and 
	M. T. Sargent
		\inst{\ref{i3}}	
		\and
	M. Brusa
		\inst{\ref{i2},\ref{i4}} 
		\and
        E. Daddi 
		\inst{\ref{i5}}	
		\and
        C. Feruglio
		\inst{\ref{i6}}	
		\and        
	G. Cresci
		\inst{\ref{i1}} 
	        \and
	G. Lanzuisi
		\inst{\ref{i2},\ref{i4}} 
               \and
	E. Lusso
		\inst{\ref{i7}} 
		\and
	A. Comastri
		\inst{\ref{i4}}
		\and
	R. T. Coogan
		\inst{\ref{i3}}
                \and
	Q.  D'Amato
		\inst{\ref{i2}}
		\and
	R. Gilli
		\inst{\ref{i6}}
		\and
        E. Piconcelli
        	\inst{\ref{i8}}     
                \and
	C. Vignali
		\inst{\ref{i2},\ref{i4}}
}


\institute{INAF - Osservatorio Astrofisico di Arcetri, Largo Enrico Fermi 5, I-50125 Firenze, Italy\label{i1}
       \and
        Dipartimento di Fisica e Astronomia, Universit\`a di Bologna, via Gobetti 93/2, 40129 Bologna, Italy\label{i2}
	\and
        Astronomy Centre, Department of Physics and Astronomy, University of Sussex, Brighton, BN1 9QH, UK\label{i3}
	\and
        INAF Osservatorio di Astrofisica e Scienza dello Spazio di Bologna, via Gobetti 93/3, 40129 Bologna, Italy\label{i4}
         \and
        CEA, IRFU, DAp, AIM, Universit\'e Paris-Saclay, Universit\'e Paris Diderot, Sorbonne Paris Cit\'e, CNRS, F-91191 Gif-sur-Yvette, France\label{i5}
        \and
        INAF Osservatorio Astronomico di Trieste, Via G.B. Tiepolo 11, I-34143 Trieste, Italy\label{i6}
        \and
         Centre for Extragalactic Astronomy, Department of Physics, Durham University, South Road, Durham, DH1 3LE, UK\label{i7}
         \and
         INAF - Osservatorio Astronomico di Roma, via Frascati 33, 00078 Monte Porzio Catone (RM) Italy\label{i8}
}

\date{Received 2 November 1992 / Accepted 7 January 1993}

\abstract {} 
{The standard active galactic nuclei (AGN)-galaxy co-evolutionary scenario predicts a phase of deeply ``buried'' supermassive black hole growth coexisting with a starburst (SB) before feedback phenomena deplete the cold molecular gas reservoir of the galaxy and an optically luminous quasar (QSO) is revealed (``SB-QSO evolutionary sequence''). 
The aim of this work is to measure the cold gas reservoir of three highly obscured QSOs to test if  their gas fraction is similar to that of submillimeter galaxies (SMGs), as expected by some models, and place these measurements in the context of the SB-QSO framework.
} 
{We target CO(1-0) transition in BzK4892, a Compton Thick (CT)  QSO at z=2.6, 
CO(1-0) in BzK8608 and CO(2-1) in CDF153, two highly obscured ($N_H\approx 6\times 10^{23}$ cm$^{-2}$) QSOs at z $=2.5$ and z $=1.5$, respectively. For all these targets, we place 3$\sigma$ upper limits on the CO lines, with $L'_{CO} < (1.5\div 2.8)\times 10^{10}$ K km/s pc$^2$. We also compare the molecular gas conditions of our targets with those of other systems at z $> 1$, considering normal star forming galaxies and SMGs, unobscured and obscured AGN from the literature. For the AGN samples, we  provide an updated and (almost) complete collection of targets with CO follow-up at z $> 1$.} 
{BzK4892 displays a high star formation efficiency (SFE $=L_{IR}/L'_{CO}> 410$ L$_{\odot}$/(K km s$^{-1}$ pc$^2$ )) and a gas fraction $f_{gas}=M_{gas}/(M_{star}+M_{gas})<10\%$. Less stringent constraints are derived for the other two targets ($f_{gas}<0.5$ and SFE $>10$ L$_{\odot}$/(K km s$^{-1}$ pc$^2$ )). From the comparison with literature data, we found that, on average, {\it i}) obscured AGN at z $>1$ are associated with higher SFE and lower $f_{gas}$ with respect to normal star forming galaxies and SMGs; {\it ii}) mildly and  highly obscured active galaxies have comparable gas fractions;  {\it iii})  the SFE of CT and obscured AGN are similar to those of unobscured AGN. 
} 
{
Within the SB-QSO framework, these findings could be consistent with a scenario where feedback can impact the host galaxy already from the early phases of the SB-QSO evolutionary sequence.
} 

\maketitle
\titlerunning{BzK4892} 

\section[Introduction]{Introduction}

The observed scaling relations between supermassive black hole (SMBH) and galaxy properties found in the local Universe suggest a close connection between the galaxy evolution and the growth of the central SMBH (see \citealt{Kormendy2013} for a recent review). Both theoretical and observational arguments (e.g. \citealt{Alexander2005,Hopkins2008, Menci2008}) support a picture, first presented by \citet{Sanders1988}, where a dusty, starburst-dominated system, arising from a merger, evolves into an optically luminous quasar (QSO). The QSO phase would emerge after the accreting SMBH experienced the so-called ``feedback'' phase (see e.g. \citealt{Hopkins2008}),  invoked to reproduce the host galaxy-SMBH scaling relations (but see also e.g.  \citealt{Jahnke2011}). During this feedback phase, the active galactic nucleus (AGN) releases radiative and kinetic energy in the form of powerful, outflowing winds, which may be capable of clearing the entire galaxy of dust and gas, thereby causing star formation in the host to cease  (see \citealt{Somerville2015} for a recent review). We will refer to this evolutionary picture as the starburst (SB)-QSO evolutionary sequence.

In the past decade, the properties of cold molecular gas reservoirs of galaxies have been the subject of intense investigation. This gas phase is regarded as the immediate fuel from which stars form, and can be used to test the starburst (SB)-QSO evolutionary sequence (e.g. \citealt{Kennicutt2012,Leroy2008}). 
Molecular gas studies, mostly involving observations of carbon monoxide (CO) emission lines, have been carried out in SB-dominated galaxies (e.g., ultra-luminous infrared galaxies, ULIRG, at low redshift, and submillimeter galaxies, SMGs, at higher redshift), unobscured QSOs, and in normal star-forming galaxies (SFGs; e.g., \citealt{Carilli2013, Daddi2010, Tacconi2013,Aravena2014}). 
These studies showed that optically luminous QSOs are characterised by low molecular gas content with respect to their current star formation rate (SFR), or alternatively by high star formation efficiency (SFE), defined as the rate of star formation per unit of molecular gas mass, SFE $=$ SFR/M$_{gas}$ or, SFE = L$_{IR}$/L$'_{CO}$\footnote{In this paper, we use the ``empirical'' SFE definition, SFE = L$_{IR}$/L$'_{CO}$, rather than the ``physical'' SFE with units of yr$^{-1}$. Here,  L$_{IR}$ refers to the $8-1000\mu$m integrated stellar infrared luminosity, while L$'_{CO}$ indicates the CO(1-0) luminosity. The advantage of using these empirical measurements is that they are not subject to the uncertainties related to the use of conversion factors needed in the ``physical'' formalism for the derivation of gas masses (Sect. 4.5 of \citealt{Carilli2013}).}, in the range between a few 100 L$_{\odot}$/(K km s$^{-1}$ pc$^2$ ) (units omitted from now on for simplicity) and $\approx 1000$ (see e.g. Fig. 4 in \citealt{Feruglio2014} for an early compilation).  On the other hand, normal star-forming galaxies and SBs display SFE  values range from $\sim 10$ and 200, with the latter population generally lying on the upper end of this range (\citealt{Feruglio2014}).
We note that the SFE of individual classes of sources is roughly constant with redshift at z $>1$ (e.g. \citealt{Sargent2014,Feruglio2014,Schinnerer2016,Tacconi2018,Yang2017}). Therefore, the characteristic SFE-ranges for optically luminous QSOs and star-forming galaxies quoted above are effectively applicable at all z>1 (see \S \ref{results1} for more detailed discussion).

\begin{figure}
\centering
\includegraphics[width=9cm,trim=0 0 0 0,clip]{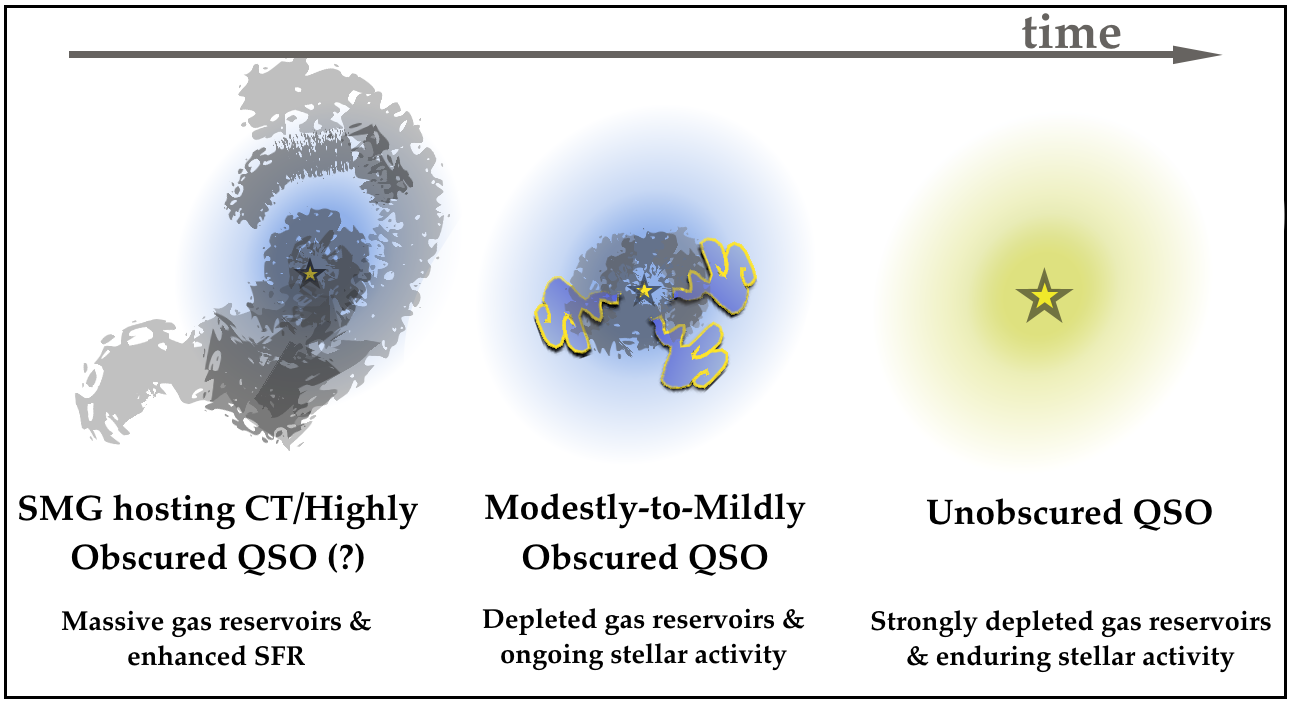}
\caption{\small 
Schematic diagram to illustrate the ``standard'' SB-QSO evolutionary sequence. In the first two panels, the blue ellipse and yellow star represent the galaxy and the central SMBH, respectively;  light and dark gray areas refer to inflowing gas and dust. Dark-blue clouds with yellow outlines indicate outflows. In the last panel, the yellow star and ellipse represent the SMBH and the QSO emission. In the first phase (left panel), the central regions are fuelled by huge amounts of gas and dust, and the SMBH is highly obscured. These systems, originating from gas-rich galaxy mergers, are commonly identified as SMGs, and are associated with star formation efficiencies of $\sim 100\div 200$. The SFE of CT/highly obscured QSOs is mostly unknown. These systems, if tracing the same evolutionary stage, should have SFEs similar to those of SMGs. In the last phase (right panel), the system is revealed as an unobscured QSO (bright enough to easily outshine its host galaxy). In this phase, its gas reservoir is expected to be strongly depleted. The SFE measured in optically luminous QSOs can span a wide range, from a few 100 to $\sim$ 2000. In the transition phase (central panel), galaxy-wide outflows start to remove the dust and gas reservoirs (which only partially obscure the nuclear regions). We observe these systems as obscured QSOs (generally associated with signatures of outflows). For these systems, we naively expect to observe intermediate SFE between SMGs and unobscured QSOs. 
}
\label{cartoon}
\end{figure}

According to the SB-QSO paradigm, 
the SFE of unobscured QSOs should be significantly higher than for main sequence (MS; \citealt{Speagle2014}) galaxies and SBs because of the AGN feedback, which expelled the gas from their host galaxies during previous phases. In particular, during the initial stages of the merger, the SFR is expected to be a factor of a few (up to 10) higher with respect to normal, isolated galaxies (\citealt{DiMatteo2005,DiMatteo2007,Hopkins2012}). This translates to an enhancement of the SFE by the same amount (\citealt{DiMatteo2007}).  An even higher SFE could be associated with the subsequent phases - i.e. the feedback phase and, in particular, the QSO phase - because the CO luminosity reflects the current (depleted) amount of molecular gas, while the IR emission responds to the associated decrease in SF more slowly than the outflow time-scales (see \S \ref{observations}; Fig. \ref{cartoon}).

In the SB-QSO evolutionary sequence, obscured AGN are associated with a transitioning phase between the SB- and QSO-dominated stages, so that the higher the obscuration, the higher the gas reservoirs in the host galaxy. In this framework, the obscuration is due to galaxy-scale absorbers rather than a nuclear dusty torus on the line of sight. Although this argument is still debated and might not be applied to all the obscured systems, there are some observational evidence confirming the presence of galaxy-scale absorbers in obscured QSOs (see e.g. \citealt{Lanzuisi2017} and references therein; \citealt{Gallerani2017,Gilli2010,Gilli2014,Rosario2018,Symeonidis2017}). This is suggested, for example, by the correlation between the nuclear column density ($N_{\rm H}$) and the amount of gas traced by the CO on kpc-scales (e.g. \citealt{Rosario2018}), and by the fact that in a few high-z CT AGN the absorber can be fully accounted for by the amount of CO gas and its compactness (e.g. \citealt{Gilli2014}). 
Moreover, numerical simulations by \citet{Hopkins2016} also show that typical nuclear conditions of post-merger objects are generally associated with isotropically obscured SMBHs with log($N_{\rm H}$/cm$^{-2}$)$\approx 24-26$. Only when the AGN-driven winds start, a cavity is generated in the direction perpendicular to SMBH accretion disk and the nuclear regions can be observed as a mildly-to-modestly obscured AGN (see also e.g. \citealt{Hopkins2005}).

The study the molecular gas conditions in different classes of sources may play and important role to test the SB-QSO evolutionary sequence. In this regard, 
scarce observations have been carried out for heavily obscured AGN  at the peak epoch of cosmic SF activity and SMBH growth (z $\sim$ 2; \citealt{Cimatti2006}). 
The identification of distant, buried QSOs is particularly challenging since the high gas column densities absorb their emission below rest frame 10 keV, so that they might be largely missed even in the deepest hard X-ray surveys available today. Because of this limitation, only a few high-z highly obscured ($N_{\rm H}> 10^{23}$ cm$^{-2}$) quasars have been targeted in CO millimeter studies (e.g. \citealt{Aravena2008, Coppin2008, Coppin2010, Polletta2011}). 

The purpose of this work is to review our understanding of the molecular gas emission in highly obscured QSO host galaxies. We present the Jansky Very Large Array (JVLA)  and the Plateau de Bure Interferometer (PdBI) observations of three obscured QSOs at z $\sim 2$. We also take advantage of the available information for z $>1$ buried systems in the literature. We test the aforementioned evolutionary scenario, i.e. whether these systems can be considered as the progenitors of optically luminous QSOs, comparing their star formation efficiencies (see Fig. \ref{cartoon}). We will also compare the cold gas reservoir of obscured AGN with those of normal MS and SB galaxies  with the same stellar mass and redshift to unveil the presence of potential feedback effects.  

The paper is organised as follows.
In Section (\S) 2 we present the targets analysed in this work, discussing their properties derived from available multiwavelength information and presenting the PdBI and JVLA observations and data analysis. \S 3 presents the samples of  normal and submillimeter galaxies, unobscured and obscured QSOs collected from the literature. In \S 4 we show the SFE and gas fractions of the different classes of sources, and discuss these properties in the context of the galaxy evolution paradigm, also taking into account their host galaxy properties. Finally, we summarise our results in the Conclusion Section.
A flat universe model with a Hubble constant of $H_0=$ 67.7 km s$^{-1}$ Mpc$^{-1}$ and $\Omega_M$= 0.307 (\citealt{Planck2015}) is adopted.  We adopt a Chabrier initial mass function to derive host galaxy properties for the targets presented in this work and for the comparison samples. 
For the X-ray detected sources reported in this paper, we distinguish between mildly obscured (with $21\lesssim$log(N$_H$/cm$^{-2}$)$<22$), modestly obscured ($22\lesssim$log(N$_H$/cm$^{-2}$)$<23$), highly obscured ($23\lesssim$log(N$_H$/cm$^{-2}$)$<24$) and CT AGN.

\begin{table*}[h]
\scriptsize
\begin{minipage}[!h]{1\linewidth}
\centering
\caption{Targets properties}
\begin{tabular}{ccccccccccccc}
 target & z&         RA & DEC     &  R$-$K &MIR/O & log(L$_X$) & log(N$_H$)& log(M$_{star}$) & log(L$_{IR}$) & SFR & log(sSFR/sSFR$_{MS}$) & SFE\\
          &    & (J2000) & (J2000) &          &          &    (erg/s)   &cm$^{-2}$   & (M$_\odot$)    &(L$_\odot$)   & (M$_\odot$/yr) & &\\
{\scriptsize (1)} & {\scriptsize (2)}  & {\scriptsize (3) } & {\scriptsize (4)}  &{\scriptsize (5)}  & {\scriptsize (6)}  & {\scriptsize (7)  }&{\scriptsize (8) } &{\scriptsize (9) } & {\scriptsize (10)}& {\scriptsize (11)}& {\scriptsize (12)} & {\scriptsize (13)}\\
\hline
BzK4892 & 2.578$^{sp}$ & 03:32:35.7 & $-27$:49:16 &  5.3 &2000 & $43.71_{-0.59}^{+0.46}$ & $24.53_{-0.20}^{+0.32}$  & $11.26\pm 0.15$ & $12.81\pm 0.01$ & $623_{-21}^{+15}$ &$0.23\pm 0.16$ &$>410$\\
BzK8608 & 2.51$^{ph}$   & 03:32:20.9 & $-27$:55:46 & 4.8  & 300  & $44.29_{-0.09}^{+0.09}$ & $23.75_{-0.20}^{+0.26}$  & $10.95\pm 0.10$      & $11.44_{-0.70}^{+0.34}$ & $27_{25}^{+32}$&$-0.88\pm 0.57$ &$>11$\\
CDF153  & 1.536$^{sp}$ & 03:32:18.3 & $-27$:50:55 &  5.3 &   70  & $44.13_{-0.04}^{+0.04}$ & $23.77_{-0.11}^{+0.12}$  & $11.11\pm 0.10$      & $11.72_{-0.07}^{+0.04}$  & $50_{-8}^{+6}$&$-0.43\pm 0.16$ &$>19$\\
\hline
\end{tabular}
\label{properties}
\end{minipage}
{\small {\bf Notes}: Column (1): galaxy name as reported in the paper. Column (2): spectroscopic redshift from literature for BzK4892 and CDF153; for BzK8608 we reported the photometric redshift derived in this paper. Column (3) and (4): equatorial coordinates. Column (5) and (6): R$-$K color in Vega system and mid-infrared (24 $\mu$m) to optical (R band) flux ratio (MIR/O) used to isolate obscured AGN in \citet{Fiore2008}. Column (7) and (8): absorption corrected 2-10 keV X-ray luminosity and column density from \citealt{Liu2017}. Column (9) to (12): host galaxy properties derived in this work (\S \ref{targets}). L$_{IR}$ referes to the $8-1000\mu$m integrated infrared luminosity; the specific star formation rate ratio is obtained as the ratio between the specific star formation rate (sSFR=SFR/M$_{star}$) and that expected for a MS galaxy at given computed stellar mass and redshift (sSFR$_{MS}$), according to the relation of \citet{Speagle2014}. The SFR measurements are obtained from IR luminosities, using the \citet{Kennicutt1998} relation. The quoted uncertainties refer to 1$\sigma$ confidence levels. Column (13): $3\sigma$ upper limit star formation efficiency, in units of L$_{\odot}$/(K km s$^{-1}$ pc$^2$ ).}
\end{table*}

\section{The targets}\label{targets}

\begin{figure*}[h]
\centering
\includegraphics[width=7.4cm,trim=20 2 60 0,clip]{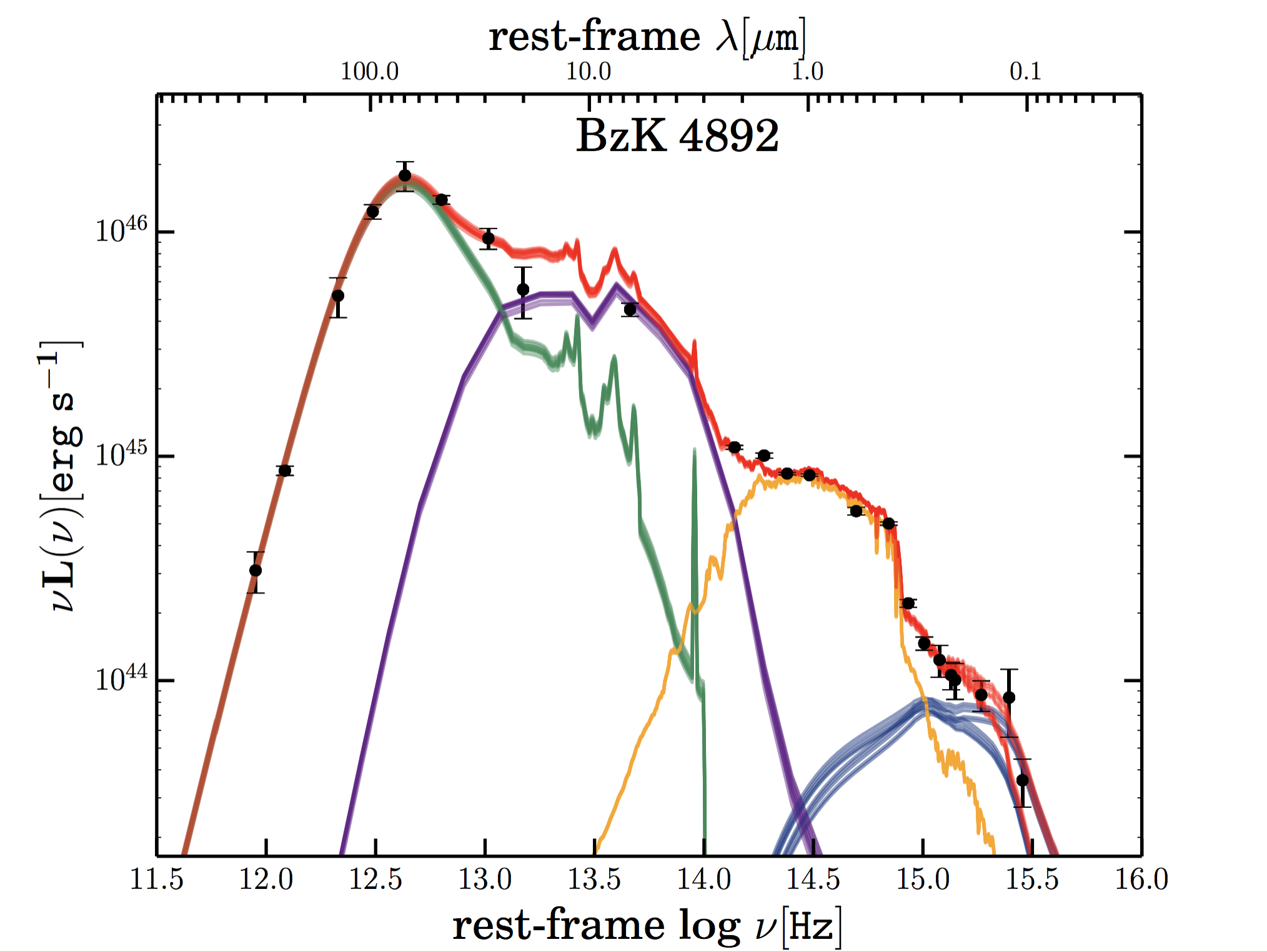}
\includegraphics[width=7.4cm,trim=10 0 30 0,clip]{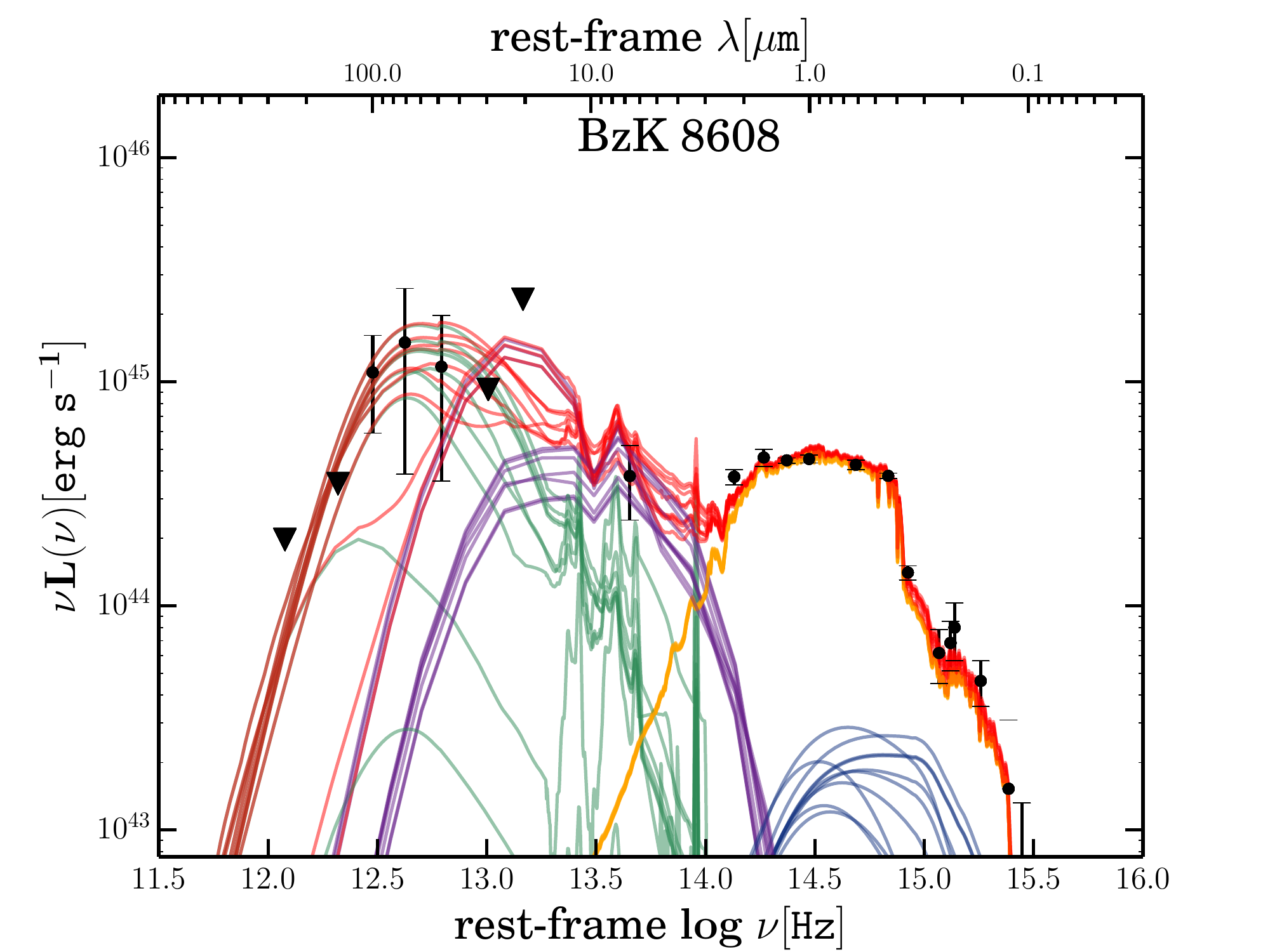}
\includegraphics[width=7.4cm,trim=10 0 30 0,clip]{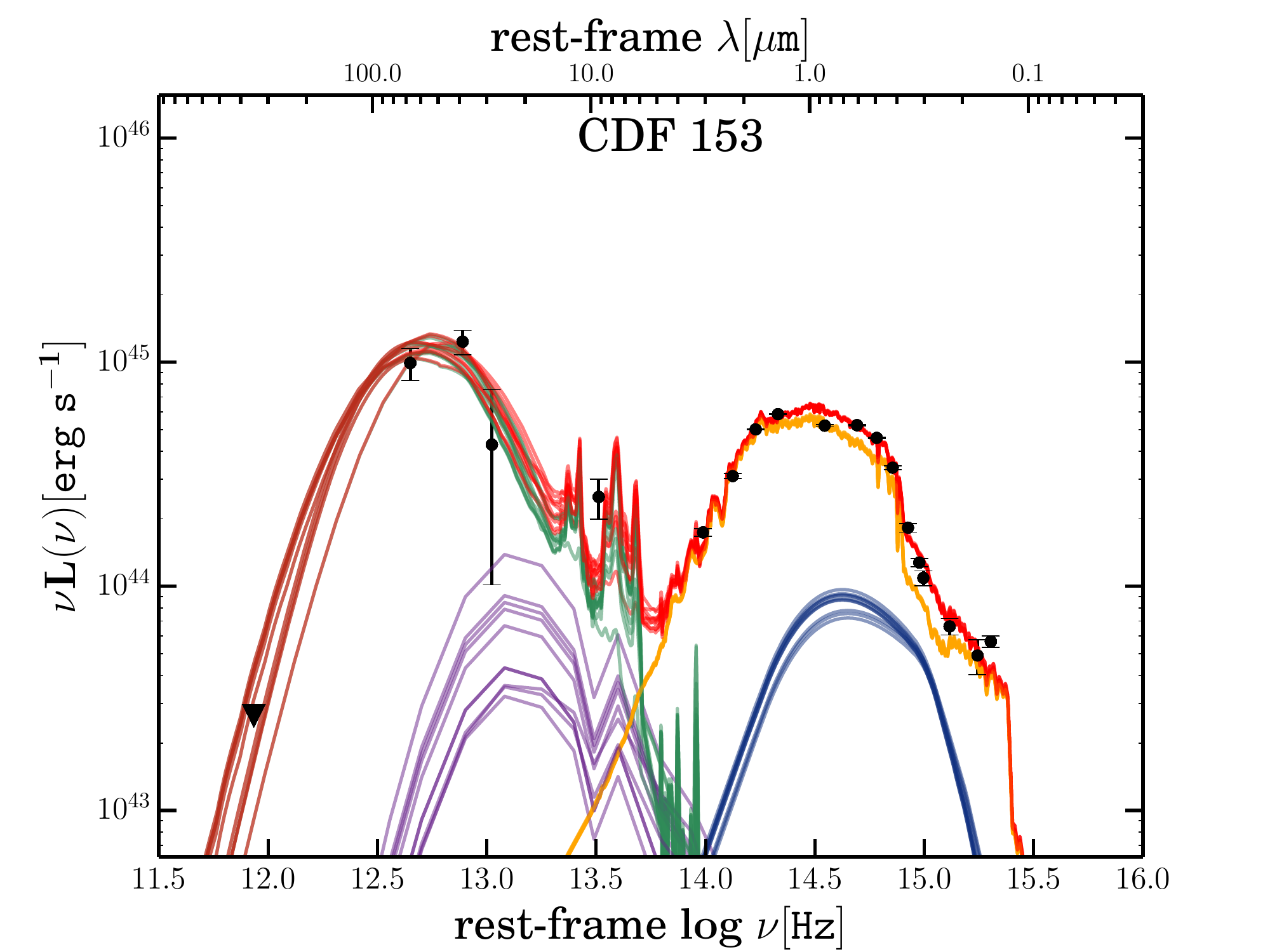}
\includegraphics[width=7.4cm,trim=0 0 0 0,clip]{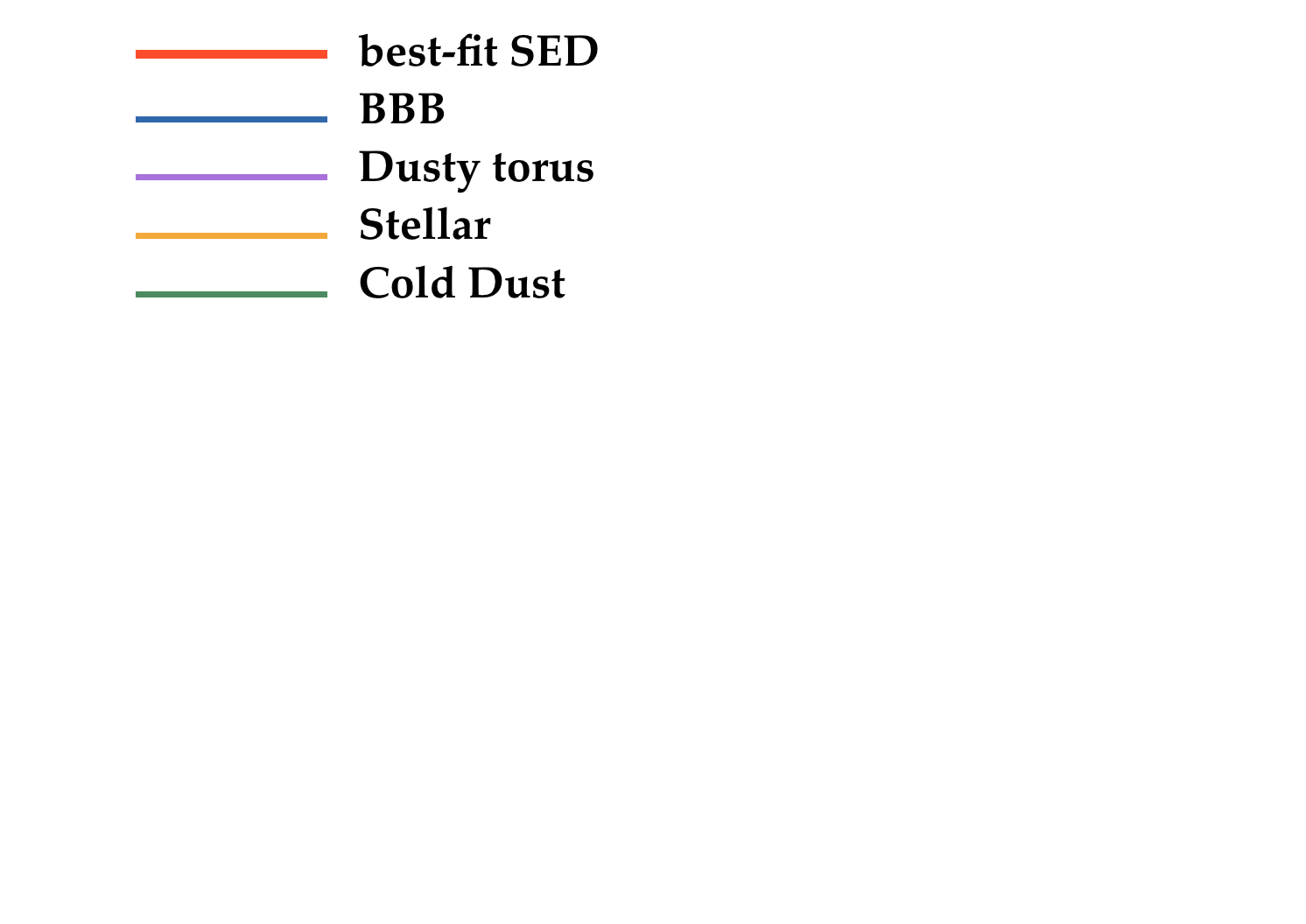}
\caption{\small SED fitting of BzK4892 (top left panel), BzK8608 (top right) and CDF153 (bottom left). The black dots represent the observed data points, collected from the \citet{Guo2013} and Wang et al. (in prep.) catalogs; $5\sigma$ upper limits are shown with black triangles. The data points at log($\nu$)$\sim 12$ are from ALMA continuum observations at 250 GHz and 340 GHz (see \S \ref{targets}). The best-fit SED obtained with AGNfitter (\citealt{Calistro2016}) and the different components used to model the SED are shown as labeled in the figure. Ten randomly picked realisations from the posterior distribution function are over-plotted in the figures, in order to visualize the dynamic range of the parameter values. }
\label{SED4892}
\end{figure*}

The identification and classification of highly obscured and CT AGN are a challenging task. In fact,  X-ray spectroscopic analysis, the most robust method of identifying these systems, is only feasible for a small number of sources because it requires high quality spectra. These sources have been detected using deep surveys by {\it XMM-Newton} (e.g. \citealt{Comastri2011,Lanzuisi2015}), {\it Chandra} (e.g. \citealt{Brightman2014,Liu2017}) and {\it Swift} (e.g. \citealt{Ricci2017}). The first publications from the {\it NuSTAR} survey programs are now further broadening the number of highly obscured sources in the Universe (e.g. \citealt{Brightman2015,Farrah2016,Marchesi2018,Masini2016,Masini2018,Ursini2017}).

Moreover, given the expected low spatial density of CT AGN in the 2-10 keV band, large area surveys are needed to collect sizeable samples. 
For this reason, alternative, multi-wavelength selection techniques based on high ionisation, narrow optical emission  lines  (\citealt{Vignali2010,Gilli2010}), or on the ratio between mid-infrared and optical fluxes (\citealt{Daddi2007,Eisenhardt2012,Fiore2008,Fiore2009}) have been developed in recent years. They seem to represent a promising approach for selecting sizeable samples of highly obscured/CT AGN at z $\sim 1 - 3$ (see e.g. \citealt{Perna2015b,Piconcelli2015,Vito2017}). However, they rely on indirect methods, and may still be prone to contamination by less obscured AGN or galaxies (e.g. \citealt{Donley2008}; see \citealt{Brandt2015} for a recent review).

The sources  analysed in this work were selected from the \citet[][BzK4892 and BzK8608]{Feruglio2011} and \citet[][CDF153]{Comastri2011} works, on the basis of the presence of high equivalent width FeK emission in the X-ray ($EW \sim 1\div 2$ keV), which constitutes unambiguous evidence for the presence of huge obscuring column densities along the line of sight. This distinguishes them from other samples of obscured AGN that have been studied in the literature, which were selected using multi-wavelength selection techniques (e.g. \citealt{Banerji2017,Brusa2017}) or the X-ray hardness ratio (e.g. \citealt{Polletta2011}). The properties of the three CO follow-up targets analysed in this work are reported in Table \ref{properties}.

Our targets have also been observed as part of the 7 Ms Chandra Deep Field-South Survey (CDFS; \citealt{Liu2017}). The new X-ray spectral analysis\footnote{ At the time of observations, X-ray properties were derived from the 4 Ms {\it Chandra} survey (for the BzK sources; \citealt{Feruglio2011}) and the ultra-deep ($\approx 3.3$ Ms) {\it XMM-Newton} survey (CDF153; \citealt{Comastri2011}). } confirmed the presence of strong iron lines in two out of three sources (BzK4892 and CDF153; see Table 4 in Liu et al.), and their high obscuration, with column density of $N_H\approx 6\times 10^{23}$ cm$^{-2}$, for BzK8608 and CDF153, and of $N_H\approx 3\times 10^{24}$ cm$^{-2}$, for BzK4892. 
We note that the CT QSO BzK4892 would be selected also by the \citet{Fiore2008} selection criteria as a candidate CT source (see Tab. \ref{properties}).

\subsection{Host galaxy properties from SED fitting}\label{SEDsect}

In order to derive the host galaxy properties of the three targets, we collected the photometric information from the UV-to-mid infrared catalog in the CANDELS/GOODS-S field, published in \citet{Guo2013}, and from a new Herschel catalog (Wang et al. in prep), which utilises an optimised algorithm  for photometry in crowded Herschel maps (see Appendix \ref{FIRphotometry}). ALMA continuum observations at 250 GHz (\citealt{Ueda2017}) and 340 GHz (\citealt{Elbaz2017}) are also used to better constrain the mm emission of BzK4892. ALMA band 7 observations obtained  as part of the programs 2013.1.00884.S (PI: Alexander D.) and 2012.1.00869.S (PI: Mullaney J.~R.) are used instead to derive $3\sigma$ upper limits on the host galaxy dust emission at 340 GHz for BzK8608 and CDF153, respectively. As the listed uncertainties on broad band photometry in published catalogues are usually underestimated (i.e. only telescope uncertainties are usually quoted), to take into account systematic errors due to e.g. confusion limit, calibrations issues (e.g. \citealt{Nguyen2010}), we increase the nominal uncertainties by a factor of 2 (see also e.g. \citealt{Shangguan2018}).  

To model the observed spectral energy distribution (SEDs), we adopted the publicly available {\it AGNfitter} algorithm (\citealt{Calistro2016}), which implements a Bayesian MCMC method to disentangle the different physical components contributing to the observed SED. The data points have been fitted with a combination of (1) a stellar component, to account for  host galaxy contribution, (2) two black hole components, referred to as ``(mid-IR) torus'' and ``big-blue bump'' (BBB), and (3) a far-IR to sub-mm emission associated with dust-obscured star formation activity in the host galaxy. 

We refer the interested reader to \citet{Calistro2016} for details on the modelling of the main components described above and the description of the AGNfitter algorithm. Here, we briefly mention that the stellar emission is modelled with stellar population models of \citet{Bruzual2003} convolved with exponentially declining star formation histories (SFH), and that stellar templates can be reddened according to the \citet{Calzetti2000} reddening law.
The AGN dust emission models are taken from \citet{Silva2004} and consider different templates for unobscured/mildly obscured AGN, modestly obscured, highly obscured and CT AGN. The torus models associated with higher obscurations are characterised by more absorbed NIR emission and mild silicate absorption at 9.7$\mu$m (see Fig. 1 in \citealt{Calistro2016}). The BBB emission is modelled using a modified version of the empirical template for type 1 SDSS QSOs constructed by \citet{Richards2006b} and considering the possible extinction by applying the SMC reddening law (\citealt{Prevot1984}) to the BBB template. Both the reddening values E(B-V)$_{gal}$ and E(B-V)$_{BBB}$ range from 0 to 1. Finally, the cold dust emission is modelled using the semi-empirical starburst template libraries by \citet{Chary2001} and \citet{Dale2002}. 

Spectroscopic redshifts are available - and used as input for the SED decomposition with AGNfitter - for BzK4892 and CDF153. They are based on VLT/FORS follow-up (\citealt{Szokoly2004}). In the case of BzK8608 we adopt a photometric redshift value of z$_{phot}=2.51_{-0.06}^{+0.16}$ (Appendix \ref{zBzK8608}).

The SED fitting was performed imposing log(N$_H$) $> 23$, based on the available information from the X-ray spectroscopic analysis.
 From the best-fit models shown in Fig. \ref{SED4892}, we obtained stellar masses of log($M_{star}$/M$_\odot$) $\approx 11.0\div 11.3$, star formation rates in the range $70\div 620$ M$_\odot$/yr and total ($8-1000\mu$m) stellar infrared luminosities  L$_{IR}\approx (2\div 60)\times 10^{11}$ L$_\odot$ (see Table \ref{properties} for individual measurements). 

We note that the BzK4892 SED best fit is well reproducing the far-IR and the optical emission, from which $L_{IR}$ and $M_{star}$ measurements are derived. The presence of a BBB emission at the bluest wavelengths is instead not well constrained and is responsible for the 0.15 dex uncertainty in the stellar mass estimate. The stellar mass of BzK8608 and CDF153 are instead better constrained (with $\sim 0.1$ dex uncertainties); on the other hand, their IR luminosities are associated with larger uncertainties (a few $\sim 0.1$ dex), being these sources much fainter and more affected by blending in the far-IR (Appendix \ref{FIRphotometry}).

Combining the stellar mass and SFR values derived with AGNfitter, we can compare the specific SFRs (sSFR=SFR/M$_{star}$) of our tagets with typical values of main-sequence galaxies.
BzK4892 lies slightly above the MS in the SFR-M$_{star}$ plane at z$\sim 2.5$ (\citealt{Speagle2014}) and can be classified as a normal MS galaxy (see Table \ref{properties}; in this work, we classified as starburst all the sources which sSFR is 0.6 dex above the MS).  

High resolution continuum images at 870 $\mu$m obtained with ALMA showed that the dust in BzK4892 host galaxy is highly concentrated, i.e. confined inside the galaxy extent seen in the HST rest frame V band (\citealt{Barro2016,Elbaz2017}, with a V band effective radius of $\sim 0.5-0.8$ kpc). Coupling the information on the spatial extent observed in the dust continuum image and the SFR inferred from the SED fit, we derive a star formation rate surface density $\Sigma_{SFR}\sim 140$ M$_\odot$ yr$^{-1}$ kpc$^{-2}$, similar to those observed in other SMGs (see e.g. \citealt{Gilli2014, Yang2017}). Therefore, BzK4892 is plausibly experiencing a very efficient and compact mode of star formation, as expected for merger-driven SB systems (e.g. \citealt{Hopkins2006}). 

BzK8608 and CDF153 have specific star formation rates compatible (within the errors) with the values expected for MS galaxies at their stellar mass and redshift, according to the relation of \citet{Speagle2014}, and are therefore associated with MS host galaxies.

\subsection{Millimeter observations}\label{observations}
BzK4892 and BzK8608 were observed with JVLA (10-12 January 2012), in the DnC configuration, as part of the VLA/11B-060 program (PI: Daddi E.).  Observations were carried out with receivers tuned to $\sim$ 32.5 GHz, corresponding to the expected frequency of the redshifted CO(1-0) emission line, for a total time on source of 4.4 hrs.  An additional window at $\sim$ 28.3 GHz was observed, in order to attempt a continuum detection in combination with line-free channels in the upper spectral window. Both spectral windows cover a bandwidth of 1 GHz each.
We reduced the data with the CASA VLA Calibration pipeline. 

CDF153 was observed with PdBI in 2012, with the array set to the C configuration, as part of the program W041 (PI. Feruglio C.), with receivers tuned to the frequency of 91.1 GHz, expected for the CO(2-1) line. The spectral set-up covered a total bandwidth of 2 GHz. PdBI data were calibrated and imaged with the GILDAS CLIC and MAPPING\footnote{\url:http://www.iram.fr/IRAMFR/GILDAS} software. The on-source time was 4.8 hours (6-antennae equivalent).

The QSO J0329$-$2357 (for BzK sources) and 0338-214 (CDF153) have been used as phase calibration sources. We used 3C 48 (for BzK sources) and MWC349 (for CDF153) for absolute flux calibrations, which yield an absolute flux accuracy of $\sim 10\%$.  The clean beam of the VLA and PdBI observations are reported in Table \ref{observations}. Note that, due to the low declination of CDF153, the associated PdBI beam is highly elongated.

After flagging of poor visibilities, the reduced data sets reached noise levels of 0.04  mJy/beam per 40 MHz channel for BzK4892, 0.03 mJy/beam per 40 MHz for BzK8608 and 0.29 mJy/beam per 120 MHz channels for CDF153 (corresponding to $\approx$ 400 km/s, for a ``natural'' imaging scheme).  

\begin{table}
\footnotesize
\begin{minipage}[!h]{1\linewidth}
\centering
\caption{JVLA and PdBI observations}
\begin{tabular}{lccccc}
target  &Freq. & Conf & On source  & FOV & beam\\
 & (GHz)& & time (hrs) & (arcsec) & (arcsec)\\
(1) & (2)  & (3)  & (4)  & (5)  & (6)  \\
\hline
BzK4892 & 32.2 & DnC &4.4 &190$\times$190& 2.66$\times$2.35\\
BzK8608 & 32.8 & DnC & 4.2 &190$\times$190 & 2.52$\times$2.10 \\
CDF153 & 91.1 & C & 4.8 & 55$\times 55$ &24$\times 5$\\
\hline
\end{tabular}
\label{observations}
\end{minipage}
{\small {\it Note.} PdBI observations for CDF153; JVLA observations for BzK targets.}
\end{table}

These sensitivities were not sufficient to detect CO line or  continuum emission in our targets. However, we could infer a stringent $3\sigma$ upper limit to the L$'_{CO}$ of $1.6\times 10^{10}$ K km/s pc$^{2}$  for BzK4892, following \citet{Carilli2013}, and considering typical a AGN FWHM$_{CO}$ of 400 km/s (e.g. \citealt{Carilli2013}; see also \citealt{Decarli2018}). This upper limit is also consistent with the CO(4-3) luminosity derived from ALMA observations,  L$'_{CO(4-3)}=(1.3\pm 0.2)\times 10^{10}$ K km/s pc$^{2}$, which will be presented in an upcoming paper (D'Amato et al., in prep.). From the  L$'_{CO}$ upper limit, we derive a gas fraction $f_{gas}=M_{gas}/(M_{star}+M_{gas}) < 10\%$, assuming that $M_{gas}=\alpha_{CO} L'_{CO}=0.8 L'_{CO}$ (e.g. \citealt{Bothwell2013,Carilli2013}) and that the interstellar medium (ISM) is dominated by molecular gas. 
We note that the upper limit on the gas fraction we derived from the JVLA observations of the CO(1-0) line is a factor of 4 lower than the values reported by \citet{Elbaz2017} and \citet{Barro2016}, derived from dust continuum emission assuming dust-to-gas and mass-metallicity relations (\citealt{Magdis2012,Scoville2016}). 
We suggest that the discrepancy between the different estimates of $f_{gas}$ is due to the large uncertainties still affecting the derivation of gas mass from dust continuum observations.  In fact, our CO-based estimate of the molecular gas mass is not dependent on any excitation ratio correction, while the luminosity-to-gas-mass conversion factor $\alpha_{CO}=0.8$ has been chosen on the basis of the remarkable compactness of the dust continuum emission in the system (\citealt{Barro2016,Elbaz2017}).  On the other hand, any dust-based estimate of $M_{gas}$ will strongly depend on the assumed metallicity and dust-to-gas ratio.

An alternative interpretation is given by the fact that CO- and dust- based molecular mass measurements can be associated with different timescales: while the CO luminosity is related  to the current amount of molecular gas in the host, the production of dust is linked to stellar processes acting over timescales of the order of $\sim 100$ Myr (e.g. \citealt{Gall2011}). Therefore, the discrepancy between the two estimates can be also explained assuming a rapid decrease of the gas reservoirs in the host galaxy, i.e. assuming recent/ongoing outflow episodes.

For the other two targets, the derived upper limits on the CO(1-0) luminosity are L$'_{CO}<2.8\times 10^{10}$ K km/s pc$^{2}$ for CDF153, assuming a 0.8 excitation correction CO(2-1)/CO(1-0), and   L$'_{CO}<1.5\times 10^{10}$ K km/s pc$^{2}$ for BxK8608. Both sources are associated with gas fractions $f_{gas}\lesssim 50\%$.  Despite the presence of an AGN, for these two targets, we conservatively assumed  $\alpha_{CO} =  3. 6$, the common value adopted for MS galaxies (e.g. \citealt{Carilli2013}). The constraints on the gas fractions decrease to $\lesssim 20\%$ if  $\alpha_{CO}=0.8$ is assumed instead (see e.g. \citealt{Brusa2017,Popping2017} for AGN-hosting MS galaxies with $\alpha_{CO}<1$). 

We note that for BzK8608, the derived quantities should be considered as tentative. In fact, given that no unambiguous spectroscopic redshift has been obtained for this target (see \citealt{Feruglio2011}; Appendix \ref{zBzK8608}), all our measurements have been inferred after adopting the photometric redshift derived in Appendix \ref{zBzK8608}. We also note that it is possible that the CO(1-0) might fall outside the receiver tuning range, which does not cover the entire 1$\sigma$ interval associated with our photometric redshift estimate (see Fig. \ref{zphot}).

\section{Literature samples}\label{compilation}

In order to explore the conditions of molecular gas in high-z galaxies and to test the SB-QSO evolutionary scenario, we combined our new observations with data from the literature, considering different classes of objects at z $>1$, from normal galaxies and SMGs to obscured and unobscured QSOs. In particular, we consider:
\begin{itemize}
\item 
a sample of 56 main sequence galaxies from the compilation of \citet{Sargent2014}, and consisting of sources presented in \citet{Daddi2010, Tacconi2013} and \citet{Magdis2012}; 
\item 
a sample of SMGs from \citet{Bothwell2013, Coppin2008, Silverman2015,Sharon2016,Yang2017} and \citet[][see below]{Yan2010}. For \citet{Bothwell2013} targets, we distinguished between SMGs with and without AGN. The \citet{Coppin2008} sample consists of 10 submillimetre-detected QSOs at z$\sim 2$, for which CO(3-2) or CO(2-1) emission lines have been detected. 
We also added the dusty highly star forming galaxies at z $>4$ collected by \citet{Fudamoto2017}, with  SFR $=2000\div 8000$ M$_\odot$/yr. The final sample of SMGs consists of 66 targets, of which 17 are associated with AGN;
\item
a sample of unobscured QSOs taken from \cite{Carilli2013}. To these, we added five recently observed targets at z$\sim 6.3$ (three from \citealt{Venemans2017}, two from \citealt{Wang2013,Wang2016}) and one source at z $\sim 2.5$ from \citet{Gullberg2016}).  We also added CO-undetected AGN from \citet{Evans1998,Guilloteau1999, Maiolino2007,Wang2011a,Wang2011} and one CO-detected target from \citet{Solomon2005}. The final sample of optically luminous AGN consists of 49 sources. 
\item
a sample of obscured AGN collected from the analysis of \citet{Kakkad2017, Banerji2017,Fan2017} and \citet{Yan2010}\footnote{The sources MIPS506 and MIPS16144 have been excluded from the sample of obscured AGN on the basis of the presence of strong PAH features in their mid-IR spectra (\citealt{Yan2010}) and, for MIPS506, the absence of AGN signatures in optical spectra (\citealt{Bauer2010}). These sources are therefore associated with the SMG sample, as per their $L_{IR}>10^{12}$ L$_\odot$.} papers. We also included further individual targets from \citet{Brusa2017,Polletta2011, Aravena2008,Spilker2016,Stefan2015,Popping2017,Vayner2017,Emonts2014}, obtaining a final sample of 36 obscured AGN.
\end{itemize}

In Appendix \S \ref{B2} we tabulate the information collected separately for each individual class of sources and provide stellar mass and MS-offset (sSFR/sSFR$_{MS}$) distributions for all the sources for which stellar masses are available. These systems are associated with massive host galaxies, with $M_{star}\approx 10^{11}$ M$_\odot$ (with a 1$\sigma$ dispersion of 0.4 dex). We note that the collected SMGs have been selected in the different papers as per their infrared luminosity $>10^{12}$ L$_\odot$ and that, for many of them, the $M_{star}$ is unknown; therefore, this sample can include an heterogeneous collection of SB and normal MS galaxies. 
In the next sections we will combine their  CO(1-0) luminosities with total IR luminosities and stellar masses to investigate the ISM conditions of the different types of galaxies, taking into account also the presence and the obscuration of accreting SMBHs.

While the ground-state transition CO(1-0) is the preferred emission line for tracing host galaxy molecular gas reservoirs, in many cases we only have access to observations of CO emission lines with $J_{up} >1$ (Table \ref{literaturesampleSMG}, \ref{literaturesample}, \ref{literaturesampleQSO}). Converting these to an estimate of the CO(1-0) luminosity requires knowledge of the spectral line energy distribution (SLED), from which it is possible to derive the luminosity ratios between CO states with $J_{up} >1$ and the CO(1-0). However, multi-J CO observations reveal a significant diversity of SLED shapes (e.g. \citealt{Bothwell2013,Brusa2017}), caused by varying ISM conditions in different systems (e.g. \citealt{Bothwell2013}); as a result, the conversion to CO(1-0) emission can be highly uncertain when the host galaxy conditions are unknown.
  
 To avoid assumptions for the excitation correction, we used the CO(1-0) measurements reported in \citet{Sharon2016} for the five sources previously observed in $J_{up}>1$ CO lines by \citet{Coppin2008} and \citet{Bothwell2013}. For all the other objects, the CO(1-0) luminosities used in this work are reported as inferred and tabulated in the different papers, therefore without uniform assumptions about the excitation correction. In fact, 
as our collected sample is based on different classes of sources, the use of a unique factor would  introduce large uncertainties.

Apart from the above mentioned issues related to the inferred CO(1-0) luminosities, caution is required here also for the ($8-1000\mu$m) infrared luminosities, since for most of the collected targets the L$_{IR}$  determination is not self-consistent. Generally, IR luminosity is obtained from SED fitting analysis, distinguishing between AGN and galaxy contributions when complete multiwavelength data are available (e.g. \citealt{Brusa2017,Kakkad2017}), or using a modified greybody model for given temperature and dust emissivity (e.g. \citealt{Banerji2017}). The reliability of the derived estimates also depends on the availability of (deep) infrared and sub-millimeter observations of individual targets (see Notes in Table \ref{literaturesample}). 

Finally, we note that a small fraction of optically luminous QSOs and SMGs are associated with lensed systems (13 SMGs from \citealt{Yang2017} and \citealt{Fudamoto2017}, 15 QSOs from \citealt{Carilli2013}). All lensed galaxy luminosities have been therefore corrected for lens magnification factors, which may be affected by large uncertainties (see e.g. \citealt{Feruglio2014,Perna2018}). 
We are confident, however, that the  L$_{IR}$ we report in this work are generally a good approximation of the dust-emission related to SF. In \S \ref{biasesSFE}, we nevertheless discuss what impact systematics errors in L$_{IR}$ measurements could have on our results, by {\it i}) considering only a subsample of sources which IR luminosities have been derived distinguishing between AGN and galaxy contributions, and  {\it ii}) exploring a scenario where IR luminosity is significantly contaminated by AGN-related emission, in keeping with \citet[][]{Duras2017}.    

\begin{figure*}
\centering
\includegraphics[width=8cm,trim=10 130 0 110,clip]{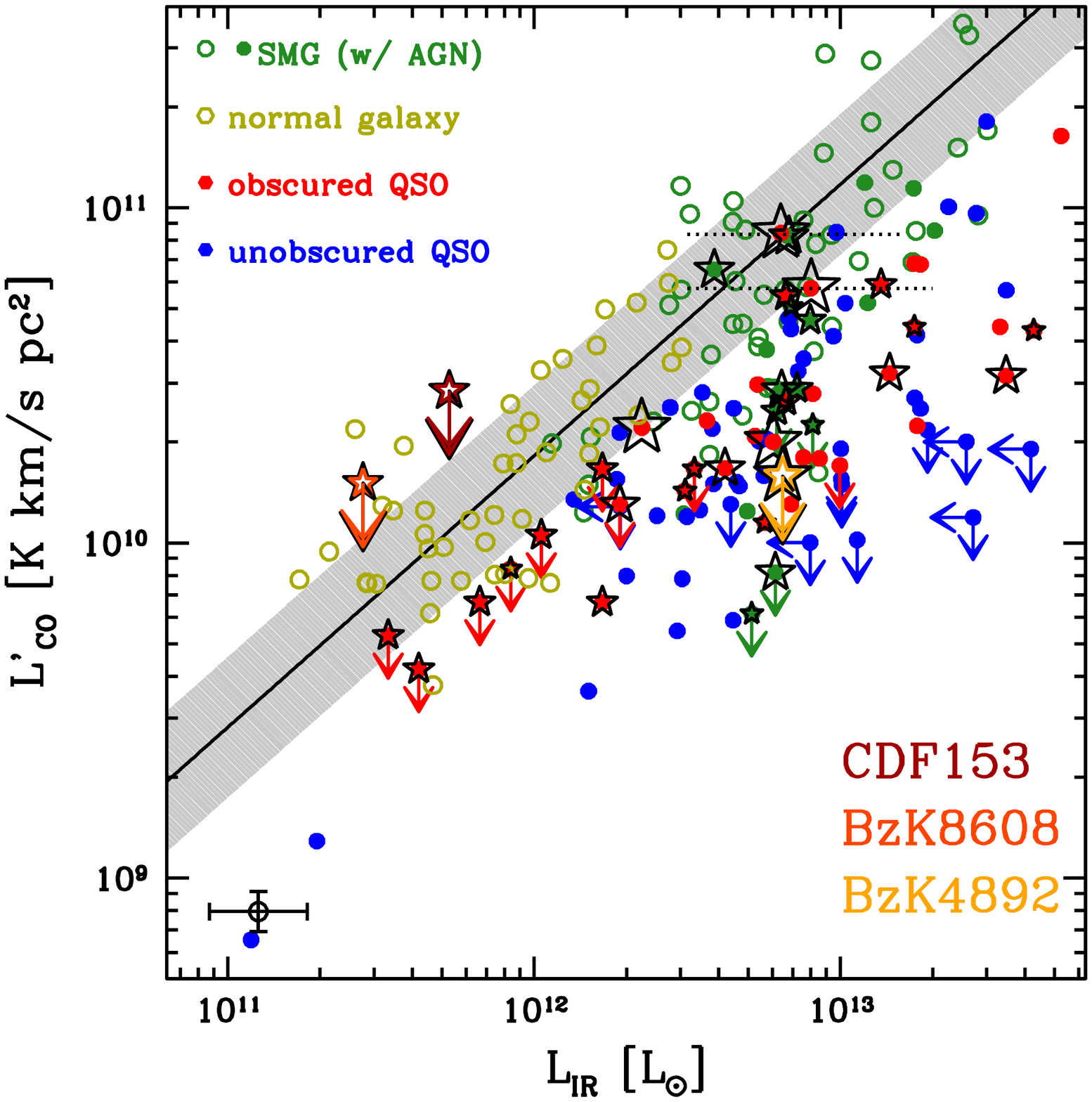}
\includegraphics[width=8cm,trim=10 130 0 110,clip]{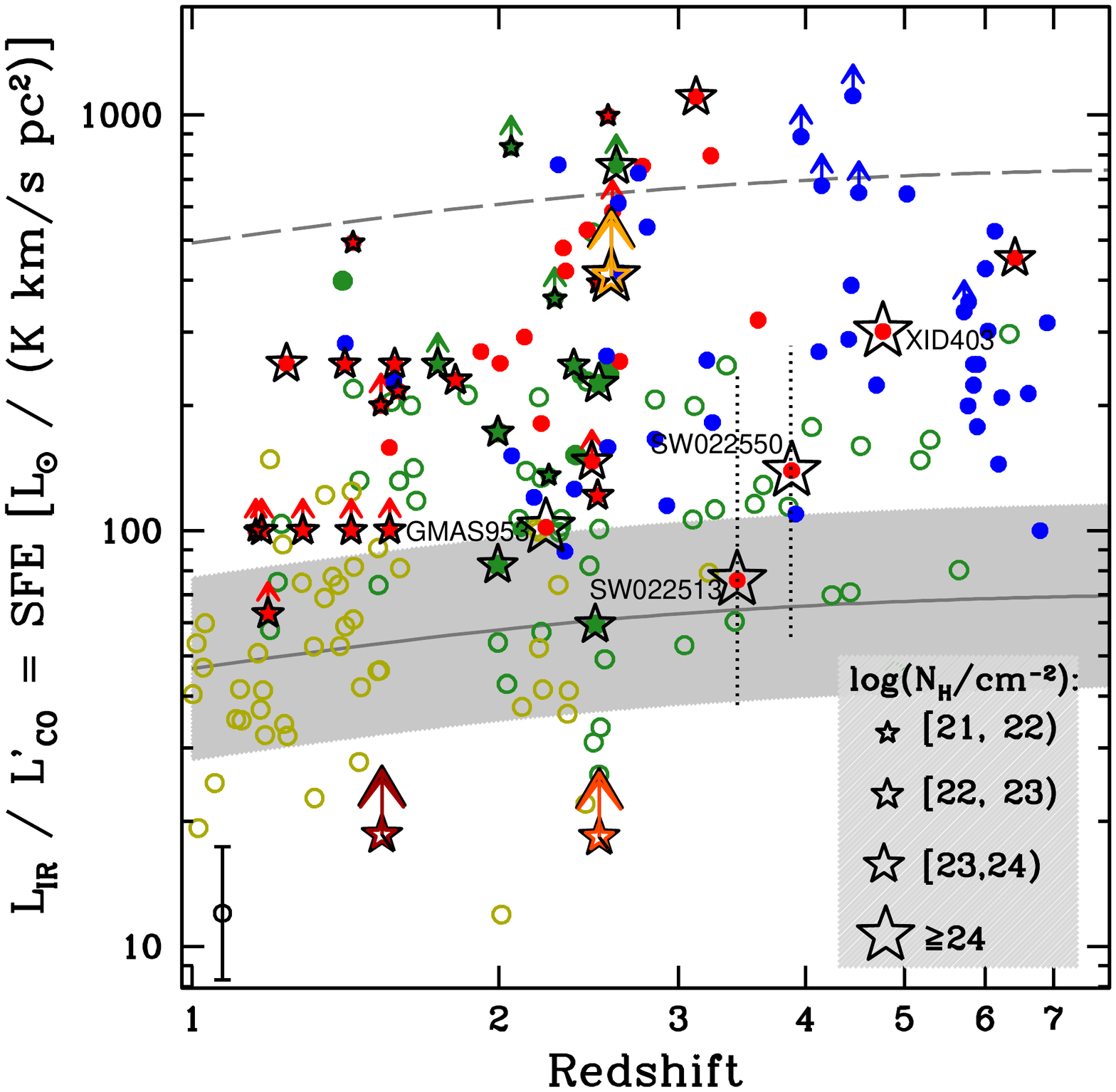}
\caption{\small {\it Left panel}: Observed correlation between measures of CO(1-0) luminosity, L$'_{CO}$, and ($8-1000\mu$m) IR luminosity, L$_{IR}$, for the samples of normal galaxies, SMGs and QSOs at z $>1$. For the three obscured QSOs described in this paper, we report the 3$\sigma$ upper limits, as labeled. Open symbols refer to main sequence galaxies and SMG, filled symbols refer to QSOs. Where column density measurements are available, scaled star symbols indicate mild to high N$_H$ values, as labeled in the right panel inset. Olive green symbol are the galaxies from \citet{Sargent2014}; green symbols refer to SMGs from \citet{Bothwell2013,Coppin2008} and \citet{Silverman2015}; red symbols refer to obscured targets. Optically bright QSOs are represented with blue dots. The solid curve represents the best-fit relation derived by \citet[][1$\sigma$ scatter indicated with the grey area]{Sargent2014}, for the MS galaxy sample they collected in the redshift range $0<$ z $\lesssim 3$. {\it Right panel}: SFE vs redshift for the same sample. In this panel, we labelled the names of the four sources from the literature associated with CT QSOs. The solid curve shows the 2-SFM predicted evolutionary trend of MS galaxies of $M_{star}=10^{11}$ M$_\odot$ (shaded area: illustration of the scatter around the mean evolutionary trend for galaxies with SFR values $1\sigma$ above/below the MS). The long-dashed curve is located at 10-fold higher SFE than the mean MS evolutionary trend. Representative $1\sigma$ errors on individual data-points are shown in the bottom-left corner of the plots. For the two CT QSOs SW022513 and SW022550, \citealt{Polletta2011} indicated a wide range for their IR luminosities (see Notes in Table \ref{literaturesample}); these sources are therefore plotted in the left panel at the centroid of the IR range, while horizontal dotted lines show the IR luminosity interval. In the right panel we similarly reported the two sources at the centroid positions and indicated the SFE ranges with vertical dotted lines.
}
\label{LcoLir}
\end{figure*}
\section{Results}

\begin{figure*}
\centering
\includegraphics[width=18cm,trim=0 0 0 0,clip]{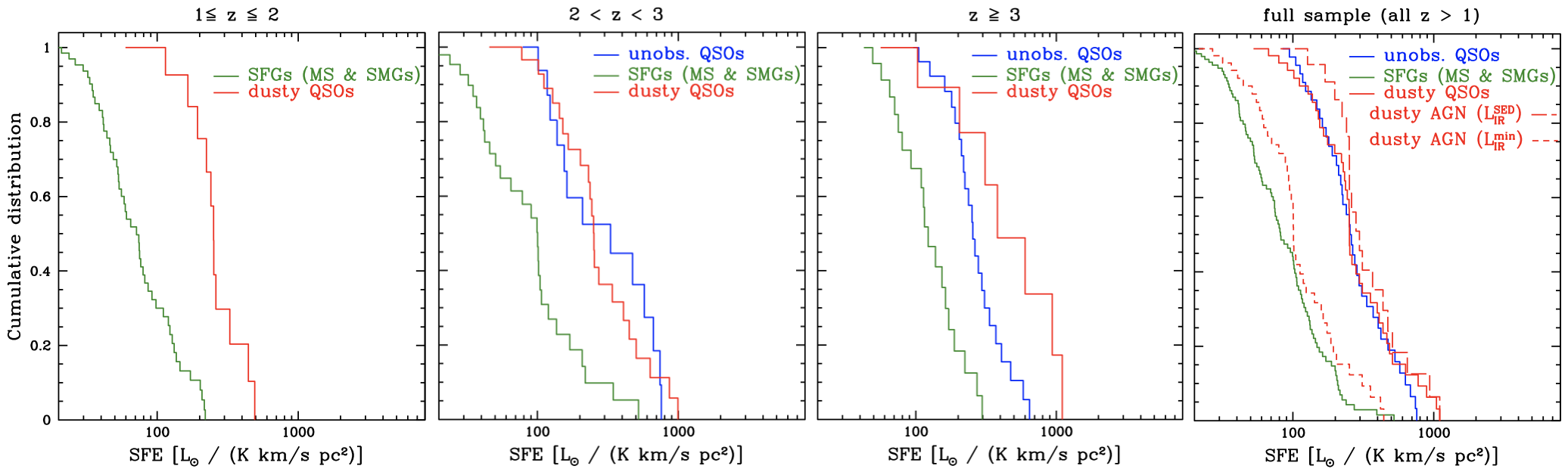}

\caption{\small  
{\it First three panels}: Cumulative distributions of SFE for the samples of star forming galaxies (green line), unobscured (blue)  and dusty (red) QSOs in the three redshift intervals specified along the upper edge of the figures. All targets collected from the literature (for which SFE measurements can be derived) are used to construct the cumulative distributions; in the first panel, we do not report the cumulative distribution of unobscured AGN (our sample only includes two of these objects at $1\le$ z $\le2$). The figures show the clear separation in SFE between AGN and SFGs. 
{\it Right panel}: Cumulative distribution of SFE for the same three samples, considering the entire redshift range covered by our targets. Short-dashed red line: cumulative SFE distribution of dusty AGN obtained by reducing by 60\% all their IR luminosities to test for the impact of cool, AGN-heated dust (e.g. \citealt{Duras2017}; see \S \ref{biasesSFE} for details). Long-dashed red line: distribution of dusty AGN for which a high-quality L$_{IR}$ has been measured through an SED-decomposition into AGN and galaxy emission.
}
\label{cumulata}
\end{figure*}

\subsection{Star formation efficiencies and AGN obscuration}\label{results1}
In Fig. \ref{LcoLir} (left) we report the CO(1-0) and infrared luminosities for all the four classes of targets mentioned above, indicating with star symbols the sources associated with mildly obscured to highly obscured and CT QSOs (the symbols are scaled so that the largest size corresponds to the highest column density; see inset of right panel). To our knowledge, the hyper-luminous galaxies hosting QSOs from \citet[][SW022550 and SW022513]{Polletta2011}, the SMG/QSO from \citet[][XID403]{Coppin2010} and the MS galaxy GMASS953 (\citealt{Popping2017,Talia2018}) are the only z $>1$ CT sources for which molecular gas has been studied\footnote{The source C92, from the sample of \citet{Kakkad2017}, previously proposed as candidate CT QSO by \citet[][XID 60053]{Bongiorno2014}, is here associated with a modestly obscured QSO, with log(N$_H$) $\sim 22.5$ cm$^{-2}$, on the basis of deeper {\it Chandra} observation (\citealt{Marchesi2016}). The source C1148, from \citet{Kakkad2017}, is instead a candidate CT source; the target is however detected with 20 net counts and deeper observations are required to confirm its high obscuration (\citealt{Marchesi2016}).}. 
The figure highlights the differences between normal galaxies, which follow the L$_{IR}-$L$'_{CO}$ relation for MS galaxies (solid line, as reported by \citealt{Sargent2014}), and the obscured QSO population, which sources are generally below the locus of MS galaxies. In general, optically luminous QSOs are further below the relation. From this plot, we note that the majority of SMGs in our sample follows the relation of MS sources.

The three obscured AGN presented in this paper are also shown in the figure. Non-stringent upper limits on the CO luminosities are associated with BzK8608 and CDF153\footnote{At the time of observations, the infrared luminosities of BzK8608 and CDF153 were overestimated of a factor of a few (as they were derived on the basis of radio 1.4 GHz observations), and the required depths in CO were settled assuming a SFE =100.}; the upper limit for BzK4892 is instead located well below the  L$_{IR}-$L$'_{CO}$ relation for MS galaxies.

In Fig. \ref{LcoLir} (right) we show the SFE as a function of redshift for the same compilation of MS galaxies, SMGs and AGN at z $>1$. Unobscured AGN show SFE in the range between 100 and 2000, while MS galaxies (SMGs) have SFE  $\lesssim$ 100 ($\lesssim$ 200). The figure also shows the obscured AGN (red symbols) and the SMGs with AGN activity (filled green symbols), with SFE in the range between $\approx 100$ and 1000. 
The figure shows no clear trend with redshift for the SFE of individual classes of targets, consistent the slow SFE-evolution at z $>1$ already discussed in the Introduction. In particular, it has been observed that the SFE of MS galaxies increases in the redshift range z $\sim 0-1$, and then flattens at z $>1$ (e.g. \citealt{Sargent2014,Santini2014}). This trend has been confirmed up to z $\lesssim 3$; at higher redshifts, CO observations of normal MS galaxies are still scarce (e.g. \citealt{Dessauges2017,Magdis2017}), and the SFE evolution is hence primarily constrained by continuum-based gas mass estimates (e.g. \citealt{Schinnerer2016}) which suggest merely slow evolution also beyond z $\sim 3$. SFE measurements of more luminous star forming galaxies, i.e. starburst ULIRGs and SMGs (with L$_{IR} > 10^{12}$ L$_\odot$), have instead been derived for large samples of targets both at low-z and in the distant Universe (out to z $\sim 6$), and, similarly to MS galaxies, do not display a strong variation with the redshift at z $> 1$ (e.g. \citealt{Aravena2016, Feruglio2014, Magdis2012, Sargent2014,Yang2017}). 
Instead, the SFE evolution of unobscured AGN population is not well known, with CO follow-up focusing mostly focused on lensed targets at z $\sim 2-3$ and on very distant (z $>4$) sources. As unobscured AGN display a wider range of SFE values than MS galaxies and SMGs, the available CO spectroscopy does not reveal any clear trends of SFE with redshift for this class of objects.

In the figure we also show the expected average evolutionary trend of MS galaxies of M$_{star}=10^{11}$ M$_\odot$ from the 2-SFM (2 star formation mode) framework of \citet{Sargent2014}. 
The 2-SFM has been introduced in \citet{Sargent2012}, and takes advantage of basic correlated observables (e.g. the MS relation and its evolution with z; \citealt{Speagle2014}) to predict the SFE (and molecular gas content; see \S \ref{gasfraction}) of MS galaxies at different redshifts.  We used a CO-to-gas conversion factor $\alpha_{CO}=3.6$ (e.g. \citealt{Daddi2010}) and the L$_{IR}-$SFR relation of \citet{Kennicutt1998} to adapt Eq. 22 from \citet{Sargent2014} to our empirical definition of SFE. We also considered the redshift evolution of the MS relation derived by \citet{Speagle2014}. 
The weak 2-SFM trend further emphasises that the redshift evolution of the SFR (and of the SFE) of MS galaxies is not responsible for the heterogenous distributions shown in Fig. \ref{LcoLir} (right) for our different  target populations.

On the other hand, SFE $>100$ values can be easily explained by assuming the presence of starburst activity, i.e. considering SFR higher than for MS galaxies (hence, enhancing $L_{IR}$), or by invoking feedback phenomena which depleted the cold gas reservoirs (hence, reducing $L'_{CO}$). As an example, we reported in the figure the predicted trend obtained by assuming a SFR ($L'_{CO}$) a factor of 10 higher (lower) than a typical MS galaxy.

\subsubsection{SFE-contrast between star-forming galaxies and AGN}\label{SFEcontrast}
As already noted by several authors, according to the SB-QSO evolutionary scenario, when an obscured QSO is detected at submm/mm wavelengths, then it could be in the transition phase from SMG to an unobscured QSO (e.g. \citealt{Coppin2008}). The high SFE values reported in Fig. \ref{LcoLir} for these sources are consistent with this scenario. In order to better visualize this result, we report in Fig. \ref{cumulata} the cumulative distributions of SFE for three different subset of our sample: (1) the MS and sub-millimetre galaxies without evidence of AGN, combined into a single sample (``SFG sample'' hereafter), as MS and SMGs are associated with similar SFE (see Fig. \ref{LcoLir}); (2) the SMGs with AGN activity (from \citealt{Bothwell2013,Coppin2008}) and the obscured AGN (hereinafter referred to as ``dusty AGN'' for simplicity); (3) the optically luminous AGN. We also divided the sources into three redshift intervals, $1<$ z $< 2$, $2<$ z $< 3$ and z $>3$, to test the presence of a possible redshift evolution in the properties of AGN targets. The cumulative distributions reported in Fig. \ref{cumulata} highlight the stark differences between star forming systems and unobscured AGN, with SFE $>200$ values for the large majority ($\sim 80\%$) of  luminous AGN  and only for a small fraction ($\sim 20\%$) of star forming galaxies. Instead, the  SFE distribution of the population of dusty AGN is similar to that of unobscured AGN.  

In order to quantitatively compare the three subsamples, we use the Kolmogorov-Smirnov test (KS test; \citealt{Press1992}). The non-parametric KS test measures the probability that two dataset are drawn from the same parent population (``null hypothesis''). The KS statistic for the three redshift intervals is $D \approx 0.6$, which corresponds to a null hypothesis probability rejected at level $\sim 90\div 99.8\%$, for both the comparisons between the SFE of SFGs and unobscured AGN and between SFGs and dusty AGN\footnote{ Lower limit values are not included in the distributions of dusty AGN and unobscured QSOs. The inclusion of these sources, representing $\sim 20\%$ of the samples of dusty AGN and unobscured QSOs, would further increase the SFE-constrast between SFGs and AGN. KS test results are obtained employing the so-called FR/RSS method (\citealt{Peterson1998}); the values reported in the text correspond to the average results from 100 iterations.}. Overall, therefore, the KS tests thus confirm the difference between star-forming systems  and AGN.
 Instead, unobscured and dusty AGN may be associated with the same parent population ($D\approx 0.4$, with a null hypothesis rejected at level $\sim 75\%$), at least at z$>2$ where we have enough unobscured AGN to perform a statistical test. 

The SFGs cumulative distributions show a clear evolution with redshift; this can be due to both the slow SFE evolution highlighted in Fig. \ref{LcoLir} and to an observational bias (at higher redshifts the sample is dominated by SMGs rather than normal galaxies). 

The medians of the SFG cumulative distributions evolve from $\sim 80$ to $\sim 130$ between the first and last redshift bin shown in Fig. 4. This is the joint effect of (a) the slow SFE evolution of MS galaxies (see Fig. \ref{LcoLir} and associated discussion in the text), and (b) an observational bias towards moderately high-SFE starburst SMGs (which represent a higher fraction of our SFG sample relative to MS galaxies at the highest redshifts). Our measurement of the significance of the SFE-gap between the AGN and SFG populations is thus a conservative estimate, as it would widen further if - at the highest redshift - our SFG sample contained similar fractions MS and SB galaxies as at z~1.
Dusty and unobscured AGN samples instead suffer from poor statistics and no clear trend can be observed.   
Given the similar SFE-offsets between AGN and SFGs at all redshifts, we constructed the total cumulative distributions of SFGs, dusty AGN and unobscured AGN in the entire redshift range covered by our literature sample (Fig. \ref{cumulata}, right). Also in this case, the KS statistic is $D \approx 0.6$ (with a null hypothesis probability rejected at level $>>99.9\%$) for both the comparisons between the SFE of SFGs and unobscured AGN and between SFGs and dusty AGN, and $D= 0.25$, with a null hypothesis rejected at level $\sim 75\%$, for the comparison between unobscured and dusty AGN. 

The position of the dusty AGN in the SFE cumulative distributions and in the SFE-z plane could suggest that their gas reservoirs have already been affected by feedback. In fact, the systems with low-to-moderate column densities could be associated with sources in which the feedback processes have already started to clean the line of sight toward the nuclear regions. This argument can be confirmed 
for the submillimetre-detected AGN in \citet[][]{Coppin2008}, for which evidence of ionised outflows have been obtained by the authors from the analysis of near-infrared spectra. Strong, multi-phase outflows have also been detected in the obscured QSO at z = 1.59, XID2028 (\citealt{Brusa2015a,Cresci2015,Perna2015a}), with SFE $\approx 250$ and atypical gas consumption conditions (i.e. very low molecular gas fraction, $f_{gas}\sim 4\%$; see \citealt{Brusa2017}).

Figure \ref{LcoLir} also show that two out of the five CT sources in our sample (BzK4892 and XID403) are strongly shifted to low values of CO luminosity (and high SFE). As discussed in the Introduction, however, we would expect such systems to lie on the L$_{IR}$-L$'_{CO}$ relation of MS galaxies (and have SFE $\lesssim 100-200$) according to the ``classical'' SB-QSO evolutionary sequence, where - at this stage - the molecular gas content should not yet have been affected by feedback according to model predictions (see Fig. \ref{cartoon}). 
Before discussing viable interpretations of these results, we take into account the possible effects of biases in the determination of the SFE values in QSO systems.

\subsubsection{Possible biases in SFE derivation for dusty AGN}\label{biasesSFE}
The empirical SFE = L$_{IR}$/L$'_{CO}$ measurements are a useful instrument for testing the model predictions, as they do not depend on the assumption to convert CO luminosity in the molecular gas mass. The SFE is also the best ``reference frame'' to compare obscured and unobscured QSOs: comparing gas masses does not provide a meaningful test, as they correlate with the mass of the systems which may differ systematically between the two samples; it is also not pratical to compare gas fractions, as optically luminous QSOs suffer from poorly constrained masses due to AGN-related emission overpowering the light from stars (this also prevents the direct comparison of cold gas properties in mass-matched samples). 

On the other hand, both the L$_{IR}$ and L$'_{CO}$  measurements (involved in estimating the SFE) are subject to systematic uncertainties. In particular, CO luminosities may be affected by excitation factor uncertainties (\S \ref{results1}). Moreover, the infrared luminosities of our sample have been derived with a variety of approaches and in some cases are poorly constrained due to non-detection in the far-IR bands (see Notes in Table \ref{literaturesample}).  In this section, we will test the impact of these biases on our results, considering the total samples of SFGs, dusty and unobscured AGN (i.e. without considering distinct redshift bins; see \S \ref{SFEcontrast}).

In order to probe whether wrong IR luminosity measurements affect our dusty AGN SFE cumulative distribution, we consider only the subsample of dusty AGN which $L_{IR}$ has been measured through an SED-decomposition into AGN and galaxy emission. Moreover, we also discarded all AGN for which IR emission is not well constrained because of the lack of data above  24$\mu$m (see Notes in Table \ref{literaturesample}). The obtained cumulative distribution is shown in Fig. \ref{cumulata}, right. A KS test demonstrated that,  even in
this case, the populations of SFGs and dusty AGN are different (D = 0.57; and null hypothesis rejected at level >> 99.9\%).

It has also been suggested that quasar-heated dust on kpc scales can contribute significantly to the far-IR luminosity (e.g. \citealt{Symeonidis2017}). This contribution, which would not be taken into account with standard SED decomposition involving a mid-IR torus component, can be as high as about 60\% for the most luminous sources (L$_{bol}\sim 10^{48}$ erg/s; \citealt{Duras2017}; see also \citealt{Symeonidis2017}). 
In order to test if the observed difference in the SFE distribution of dusty AGN and star forming galaxies is due to this bias in L$_{IR}$ measurements, we applied a correction to the SFE values by reducing the IR luminosity by a factor of $60\%$ for all the dusty AGN. We note that this is a very conservative approach:  with average absorption-corrected 2-10keV luminosities of $3\times 10^{44}$erg/s (see Table C.2; this corresponds to an average bolometric luminosity of $10^{46}$erg/s, assuming a bolometric correction of 30 following \citealt{Lusso2012}), our dusty AGN are generally much fainter than both the WISSH QSOs presented by \citet{Duras2017} and the sources discussed in \citet{Symeonidis2017}. 

The distribution derived with rescaled L$_{IR}$ is shown in Fig. \ref{cumulata}, right (red dashed line). Even by considering this extreme correction, the SFG and buried AGN cumulative distributions are significantly different (null hypothesis rejected at level $\sim 93\%$; an overlap of the two distributions only becomes possible if we assume that $>75\%$ of the IR luminosities reported in Table \ref{literaturesample} are associated with quasar emission).  We can therefore reasonably exclude that the difference between dusty AGN and SFGs cumulative distributions is due to an overestimation of the AGN host galaxy infrared luminosities.  

The excitation factors generally used to derive CO luminosities from $J_{up}>1 $ states of dusty AGN are $0.8-1.0$ (Notes in Table \ref{literaturesample}). We note that smaller values have also been proposed in the literature for the excitation corrections, in particular for high $J_{up}$ states ($J_{up}>2$; e.g. \citealt{Bothwell2013,Yang2017,Sharon2016}; up to a factor of 2 smaller); therefore, possible biases in the CO luminosity determination could be responsible for a slight shift to the left in Fig. \ref{cumulata}. We therefore consider the SFE distribution of only those sources which M$_{gas}$ is derived from CO(1-0) and CO(2-1) line fluxes, less affected by excitation factor uncertainties,  and perform a KS test between the SFGs sample and the subsample of dusty AGN associated with low J transitions. We prove that, even in this case, the two populations are different (D $=0.72$; and null hypothesis rejected at level $>> 99.9\%$). 

 To summarise, we find no evidence that the observed differences between dusty AGN and star forming galaxies SFEs are caused by observational biases.

\subsection{Low gas fractions in dusty AGN}\label{gasfraction}

\begin{figure*}
\centering
\includegraphics[width=8.3cm,trim=10 140 0 110,clip]{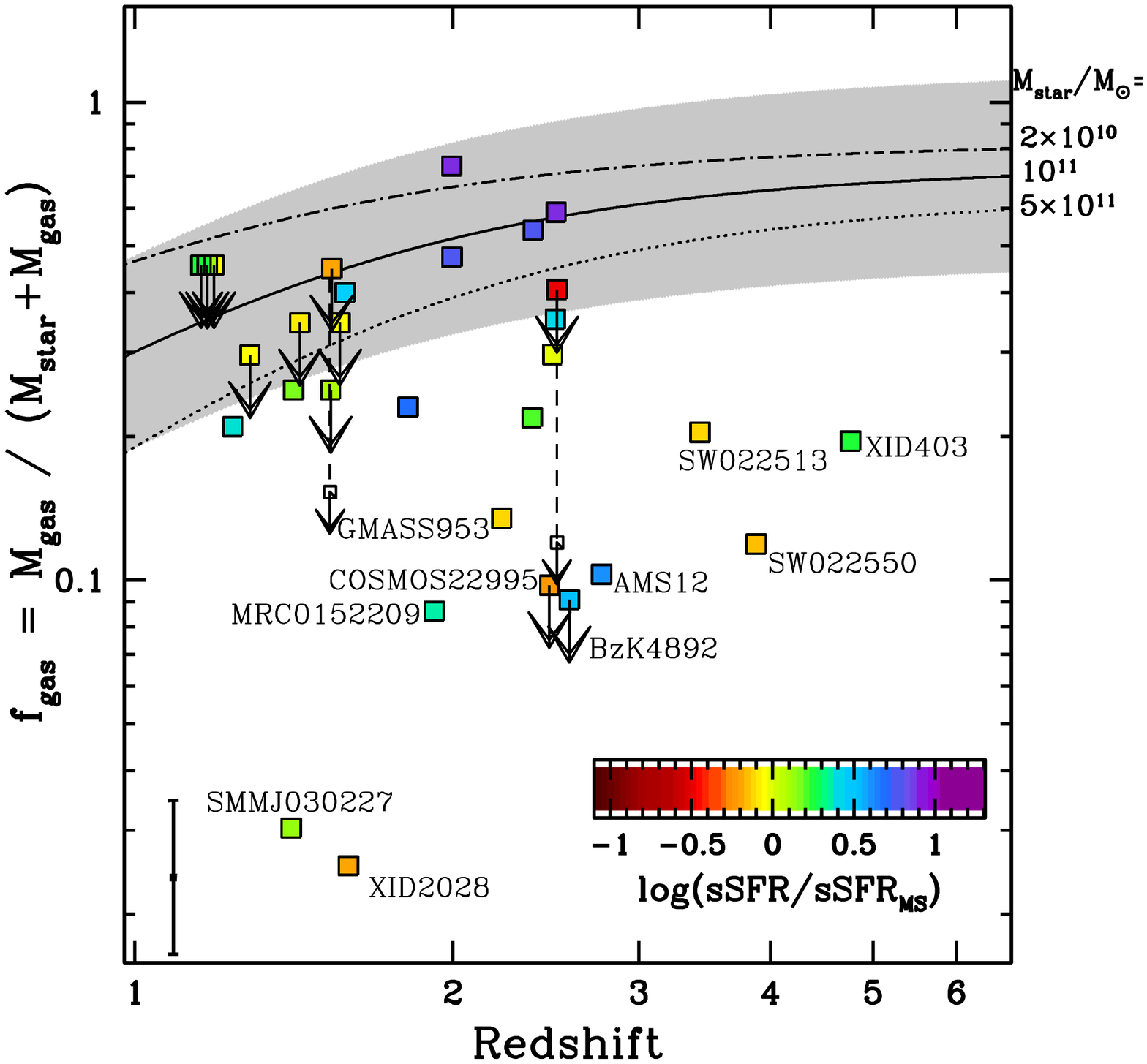}
\includegraphics[width=8.3cm,trim=10 140 0 150,clip]{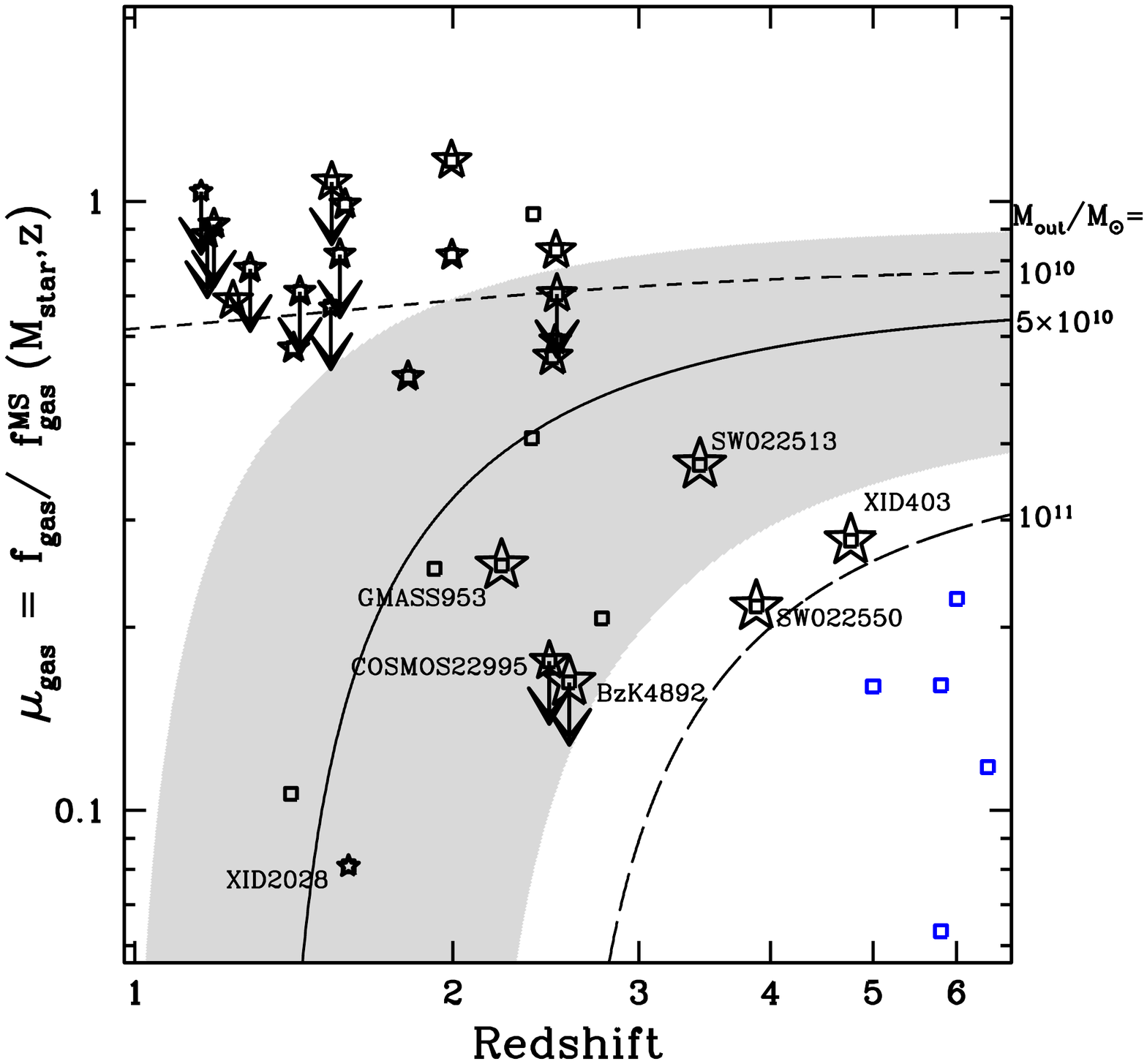}

\caption{\small  {\it Left panel:} Redshift evolution of the gas fraction $f_{gas}$ for our sample of dusty AGN. The targets are color-coded according to the distance from the main sequence of SFGs (see color bar in the bottom right). For BzK8608 and CDF153 we also show (with open symbols) the upper limits we would obtain by assuming $\alpha_{CO}=0.8$ instead of 3.6 (see \S \ref{observations}). 
The predicted evolutionary trend of MS galaxies of M$_{star}=10^{11}$ M$_{\odot}$ from the 2-SFM model is shown with a solid curve (shaded area: $1\sigma$ scatter around average evolutionary trend line). We also show the predicted trends for MS galaxies of M$_{star}=2\times 10^{10}$ M$_{\odot}$ and M$_{star}=5\times 10^{11}$, as labeled in the figure.  
The error-bar at the bottom left corner of the plot shows the representative 1$\sigma$ error for the data points (we assume a 0.2 dex uncertainty for M$_{star}$ estimates; see Table \ref{literaturesample}). 
{\it Right panel:} redshift evolution of the normalised gas fraction $\mu_{gas}$ for our sample of dusty AGN. 
The predicted evolutionary trend of ``gas depleted'' MS galaxies of M$_{star}=10^{11}$ M$_{\odot}$ is shown with the black curve (the shaded area indicates $1\sigma$ scatter around the average trend). This trend has been obtained in the framework of the 2-SFM model (\citealt{Sargent2014}) by considering the depletion of gas due to AGN-driven outflows ($M_{out}= 5\times 10^{10}$ M$_{\odot}$). We also display the predicted trends for $M_{out}= 10^{10}$ and $10^{11}$ M$_{\odot}$. The blue symbols refer instead to the predicted $\mu_{gas}$ of high-z unobscured QSOs from \citet{Valiante2014}. The star symbols are associated with X-ray detected AGN and are identical to those used in Fig. \ref{LcoLir}.  Essentially all obscured AGN hosts show a gas fraction deficit compared to normal galaxies, with CT QSOs (large star symbols) displaying among the strongest deficits.
}
\label{fgasplot}
\end{figure*}

Figure \ref{fgasplot} (left) shows the gas fractions of the 26 dusty AGN for which stellar mass estimates can be derived from multiwavelength data, as a function of redshift. 
The gas fractions of optically luminous QSOs, instead, are mostly unknown from the observational point of view, and are not reported in the figure\footnote{ 
The difficulties in deriving unobscured QSO host galaxy stellar masses from SED fitting (see \S \ref{biasesSFE})  make gas fraction estimates for these sources unreliable. An alternative approach is to infer M$_{star}$ via a dynamical mass based on molecular line observations, using the relation M$_{star} \approx$ M$_{dyn} -$ M$_{gas}$  (e.g. \citealt{Venemans2017,Wang2010}). However, these measurements are affected by very large uncertainties.}.

In Fig. \ref{fgasplot} we color-coded the sources according to their distance from the MS (from $\sim 1$ dex below to $+1.2$ dex above the sequence; see accompanying colour-bar).  Most galaxies in our SFGs sample and dusty AGN lie on the MS, with the latter population showing slightly higher MS-offsets sSFR/sSFR$_{MS}$ (see Fig. \ref{stellarmass}).

In the figure, the gas fractions are compared with the expected average evolutionary trend of MS galaxies of $M_{star}=10^{11}$ M$_\odot$ from the 2-SFM framework of \citet{Sargent2014}. The 2-SFM trend well reproduces the behaviour of massive (M$_{star}\approx 10^{10}-10^{11}$ M$_\odot$) MS galaxies (e.g. \citealt{Schinnerer2016}) and, in this mass range, is also consistent with other gas fraction evolutionary trends in the literature (e.g. \citealt{Dessauges2017,Genzel2015,Tacconi2018}).

It is still debated whether SBs are expected to have gas fractions that are similar (e.g. \citealt{Sargent2014}) or higher than for MS galaxies (e.g., \citealt{Genzel2015,Tacconi2018}). Nevertheless, the 2-SFM trend well reproduces the distribution of the SFGs presented in this work. These targets are however not reported in the figure, in order not to over-populate the plot. 

Figure \ref{fgasplot} (right) also shows that our dusty AGN are preferentially located below the gas fraction relation expected for MS galaxies. 
 Moreover, many of the AGN data points that currently lie on top/above the 2-SFM trend are upper limits. This finding is in line with the idea that dusty AGN are associated with the feedback phase, where the molecular gas content has been depleted by powerful outflows. 

Both recent cosmological simulations (e.g. \citealt{Costa2017}) and observations (\citealt{Brusa2017,Fiore2017}) suggest that AGN-driven outflows can have mass outflow rates of the order of $10^2-10^4$ M$_\odot$/yr and that the typical time-scales associated with AGN winds are of the order of $\sim 10$ Myr (e.g. \citealt{Barai2018,Costa2017,Fiore2017,Hopkins2005,Hopkins2016,Perna2015a, Valiante2014} for both observational and theoretical arguments). There is also evidence that ejected material is multi-phase, and predominantly composed of cool gas (e.g. \citealt{Costa2017,Zubovas2014}; see also \citealt{Brusa2017,Fiore2017}). 
Cosmological simulations show that the outflow masses can reach $\approx (5-80)\times 10^{9}$ M$_\odot$ during the feedback phase (\citealt{Costa2017} and references therein). Interestingly, this estimate is also consistent with observational findings: an outflow mass of $10^{10}$ M$_\odot$ can be derived assuming a typical outflow mass rate of $\dot M_{out}\approx 10^3 M_\odot$/yr and an outflow time-scale of $10$ Myr.

We consider all these arguments to derive some predictions for the gas fraction of dusty AGN associated with $M_{star}=10^{11}$ M$_\odot$ host galaxies. If we assume that dusty AGN are caught 10 Myr after the initial outflow, their gas reservoirs would be depleted by $\approx 5\times 10^{10}$ M$_\odot$; moreover, during this time a given amount of gas is converted into stars, so that $M_{gas}(t=10\ {\rm Myr})= {\rm M}_{gas}^{MS}- {\rm M}_{out} - {\rm SFR}\times \Delta t$, with $M_{gas}^{MS}$ from \citet{Sargent2014}. Assuming a constant SFR (from the $M_{star}-$SFR relationship), the term ${\rm SFR}\times \Delta t$ is however negligible with respect to ${\rm M}_{out}$. 
In Fig. \ref{fgasplot} (right) we show the predicted trend for ``gas depleted galaxies'' with the associated  1$\sigma$ confidence interval. Such large interval is mostly due to the intrinsic dispersion in the SFR-M$_{star}$ MS relation ($0.2$ dex; \citealt{Speagle2014}). We also note that at lower redshifts the outflow mass could be overestimated as the cosmological simulations are generally performed at higher redshift (z $\sim 6$), where host galaxies are associated with more rich gas reservoirs (and the AGN winds can affect larger amounts of gas; see e.g. Fig. \ref{fgasplot}, left).

In the figure, we also show the normalised gas fractions predicted by \citet{Valiante2014} for optically luminous QSOs at $5\le $ z $\le 6.4$. The authors presented a semi-analytical model for the formation and evolution of high-z QSOs and their host galaxies according to the SB-QSO scenario. They studied the evolution of the stellar and gas masses in order to explain the observed properties of $5\le $ z $\le 6.4$ QSOs. We considered the  predicted stellar and gas masses at the end of the evolutionary tracks they presented for five distinct high-z QSOs (Fig. 6 and 7 in Valiante et al.) to derive $f_{gas}$ estimates, and reported in the figure the ratios between these values and the $f_{gas}^{MS}(M_{star},z)$. Finally, depleted gas reservoirs in $z\sim 6$ galaxies experiencing AGN feedback are also found in the hydrodynamical simulations by \citet{Barai2018}, with $f_{gas}\sim 20-30\%$ lower with respect to the prelude of the feedback phase (private communication).  

An accurate comparison between different predictions and theoretical feedback prescriptions is beyond the purpose of this paper. We refer the reader to, e.g. \citet[][and references therein]{Harrison2018} for a relevant recent review. Here, we simply note that the low gas fractions (and $\mu_{gas}$) we derived for dusty AGN could be explained by recent outflows which depleted their host galaxy cold gas reservoirs.  

An alternative interpretation of our results could be that such low $\mu_{gas}$ are due to a wrong estimate of the gas mass because of a wrong assumption on $\alpha_{CO}$. In fact, a difference of a factor of 4.5 in $M_{gas}$ easily emerges when we consider $\alpha_{CO}=3.6$  instead of 0.8, the value normally used for high-z AGN and SMGs. We note however that even assuming $\alpha_{CO}=3.6$ for all the sources in our sample (i.e. SFGs and AGN), the population of dusty AGN is still associated with  lower  $\mu_{gas}$ values with respect to SFGs. Moreover, for several dusty AGN the $\alpha_{CO}=0.8$ value has been chosen on the basis of the observed compactness and/or temperature of the cold gas reservoirs in the AGN host galaxies (e.g. BzK4892, XID2028, XID403, GMASS953). 

We also note that for the few dusty AGN associated with $\alpha_{CO}=3.6$ in Fig. \ref{fgasplot}, i.e. BzK8608, CDF153 and the \citealt{Kakkad2017} targets, the CO-to-gas factor has been assumed consistently with other MS galaxies (because of their sSFR), ignoring the presence of the AGN and the possibility that a factor of 4.5 smaller amount of cold gas could be present in these systems (\citealt{Kakkad2017}).

In  Fig. \ref{fgasplot} (right) we also indicate with variably-sized star symbols the column densities associated with individual AGN (as in Fig. \ref{LcoLir}). We note that all the sources with lowest $\mu_{gas}$ ($f_{gas}\lesssim 0.2$) are associated with high obscuration (log($N_H$)$\gtrsim 23$ cm$^{-2}$), with the exception of XID2028 (log($N_H$)$\sim 21.84$ cm$^{-2}$)\footnote{\label{SMMJ}Similar gas fractions are also associated with AMS12, MRC0152-209 and SMMJ030227, for which we do not have X-ray information. HST/ACS observations of the latter target show that this SMG/QSO is a complex system surrounded by a diffuse halo of material in a ring-like structure, with a compact companion at $\sim 1.3''$ ($\sim 11$ kpc; \citealt{Swinbank2006}). Spectroscopic observations of this peculiar target also suggest the presence of complex kinematics between the nuclei (\citealt{Harrison2012,Menendez2013}).}. In particular, all CT QSOs in our sample have strongly depleted molecular gas reservoirs, with  $f_{gas}\lesssim 0.2$ and $\mu_{gas}\lesssim 0.4$.

We also derived the ratios between the (2-10) keV and IR luminosities per individual target with known, absorption-corrected X-ray  luminosity. In the framework of a simultaneous evolution between SMBH and host galaxy (\citealt{Kormendy2013}), this luminosity ratio can be used to distinguish between SF- and AGN-dominated systems, using as a reference threshold the ratio log($L_X/L_{IR}$) $=-2.2$ (see Sect. 4.4 in \citealt{Ueda2017}). We found that all but a few sources in our sample are associated with dominant AGN activity (i.e. log($L_X/L_{IR}$) $>-2.2$) and that there is  no obvious correlation between $\mu_{gas}$ and $L_X/L_{IR}$. 
The high fraction of AGN-dominated starbursting systems in the sample can be plausibly due to observational bias: the multiwavelength information required to recover stellar masses, infrared and X-ray luminosities could be, in fact, only available for the most extreme sources at z $>1$. 

Summarising, we conclude that the gas fraction of dusty AGN and, in particular, of  CT systems strongly differ from MS and SB galaxies without evidence of AGN activity.

\subsection{SB-QSO evolutionary sequence vs. orientation based unification scheme}\label{mergervsorientation}

Gas-rich galaxy major mergers have been proposed to funnel a significant amount of matter toward the nuclear regions, triggering SMBH and starburst activity. Evidence supporting a merger-driven scenario has been recently proposed by \citet{Fan2017}, who report CO line emission from a hyper luminous, dust-obscured QSO at z $\gtrsim 3$ suggesting the presence of gas-rich major mergers in the system (see also \citealt{Polletta2011}). Gas rich mergers related to high-z SMG/QSOs have also been recently presented by \citet{Trakhtenbrot2017} and \citet[][see also \citealt{Banerji2017,Vignali2018}]{Fogasy2017}.  
In the SB-QSO framework, these systems would be associated with the initial evolutionary stages, before feedback phenomena start to deplete the cold gas. 

Moreover, massive, powerful multi-phase outflows have been discovered in several local (\citealt{Feruglio2010, Perna2017a,Sturm2011,Veilleux2017}) and high redshift galaxies hosting obscured AGN (\citealt{Brusa2017,Cresci2015,Fan2017,Feruglio2017,Nesvadba2016,Popping2017}), demonstrating that luminous AGN are capable of expelling large amounts of gas from the host galaxies, thereby potentially explaining the low gas fraction (and high SFE) of optically luminous QSOs. We note however that from an observational point of view, there is not a clear separation between the two classes of dusty AGN, the first one associated with the prelude of the feedback phase (and therefore with more massive gas reservoirs and higher obscuration) and the second one associated with the feedback stage (and moderate column densities). This is the reason why CO-follow up of CT AGN, more reasonably associated with the prelude of the blow-out phase, are essential to test the SB-QSO scenario.

We find that the molecular gas reservoirs of dusty AGN are significantly different from those of SFGs (i.e. MS and submillimetre galaxies without evidence of AGN): Fig. \ref{LcoLir} and \ref{cumulata} show that SFE in the dusty AGN sample are higher than those of SFGs. This difference cannot be removed considering the biases in CO and infrared luminosity determinations. In Fig. \ref{fgasplot} we also show that the gas fraction of dusty AGN is generally lower than that expected for MS and SB galaxies.
However, the SFEs of dusty AGN and unobscured QSOs are similar (and clearly higher than those of SFGs), counter to what we would naively expect for sources in a transition phase between SBs and optically bright QSOs. 
This result could suggest that the similarly enhanced SFEs in optically bright and dusty AGN do not imply a temporal sequence,  
but that similar mechanisms are in act in these systems.

Considering only the CT QSOs, we find that two out of the five observed systems  (i.e. BzK4892 and XID403) are associated with high SFE ($\ge 300$). This finding as well is at odds with the SB-QSO evolutionary paradigm, as one would expect that the star formation efficiency of CT sources should be more similar to that of starburst galaxies (i.e. $\lesssim 200$; see Fig. \ref{cumulata}). Moreover, all the CT AGN in our sample have $\mu_{gas}\lesssim 0.4$ (Fig. \ref{fgasplot}, right).

One possible explanation for the high SFE values and low gas fractions in CT sources is that these powerful dusty AGN might be already able to quickly clear the gas from the host disk through feedback in this early stage. This hypothesis is confirmed by the presence of strong ionised outflows in the other CT QSOs in our sample, SW022550 and SW022513 (\citealt{Polletta2011}), for which deep spectroscopic information have been used to study the gas kinematics in the host galaxies, and GMASS953, for which there are several indications of the presence of AGN-driven large-scale outflows (\citealt{Talia2018}; see also \citealt{Perna2015b} for another example of powerful outflow in a CT QSO at high-z). We note however that these three CT AGN are also associated with lower SFEs with respect to XID403 and BzK4892 and, therefore, that more extreme conditions might be present in the latter AGN (e.g. the concomitant presence of strong SF and AGN activity).   

An alternative scenario is given by the possible presence of positive feedback in the host galaxies: AGN could in fact enhance the SF by outflow-induced pressure (e.g. \citealt{Cresci2015,Cresci2015b,Cresci2018,Silk2013}; see also \citealt{Maiolino2017}). In this case, the high SFR observed in the host galaxy may be associated with residual star formation activity in an already gas depleted host galaxy.  
The available information does not allow us to discriminate between positive and negative feedback scenarios; with deep, spatially resolved CO spectroscopy, it would be possible to map the molecular gas reservoirs and study the effects of outflows (e.g. \citealt{Brusa2017,Carniani2017}).

Finally it is worth noting that our results could be discussed in the context of the unified model (\citealt{Antonucci1993}). According to the orientation based unification scheme, orientation effects relative to an obscuring medium which hides the view of the inner SMBH are responsible for the differences between obscured (type 2) and unobscured (type 1) QSOs\footnote{AGN with discordant optical (type1/type2) and X-ray (unobscured/obscured) classification have been reported in the literature. In this work we do not take into account this discrepancy, since these sources should represent a small fraction of the AGN population (see e.g. \citealt{Merloni2014,Ordovas2017} for details).}. Therefore, these systems should be associated, on average, with similar host galaxy properties (i.e. molecular gas content and infrared luminosities). 
There is evidence pointing towards the fact that obscured quasars have higher average SFR and infrared luminosities than the unobscured systems (e.g. \citealt{Chen2015,Hiner2009,Zakamska2016}). 
Given this observed diversity in IR luminosity and assuming that both QSO classes have similar molecular gas reservoirs, we should observe the highest SFE in dusty systems, in contradiction with our results. 
This could be interpreted as evidence that unified model cannot explain the observed SFE of high-z QSOs.  

Further investigation is however required to probe the absence of biases in actual observations, in particular to discriminate between galaxy-scale absorbers and the presence of a torus in dusty AGN, although gas rich-galaxies could be preferentially selected within type 2 AGN samples (see e.g. the discussion in \citealt{Zakamska2016}). We also note that, under the assumption that cold gas is isotropically distributed in the host, the high column densities observed in all CT QSOs but SW022513 (for which we do not have information about CO extension), can be accounted for by the presence of cold gas on kp scale, rather than, or in addition to the nuclear (pc-scale) torus (see \citealt{Gilli2014} for details).
A further caveat is the small number of CT sources with available CO spectroscopy. In the future, a larger sample of this class of sources will be helpful in confirming our results.

\section{Conclusions}
We presented JVLA and PdBI observations for two highly obscured (BzK8608 and CDF153) and one Compton-thick (BzK4892) QSOs. We targeted the CO(1-0) line in BzK4892 and BzK8608, both at redshift z $\sim 2.5$, and the CO(2-1) transition in CDF153, at redshift z $=1.536$.  
These observations allowed us to place upper limits on the strength of the CO emission lines, corresponding to L$'_{CO} < (1.5-2.8) \times 10^{10}$ K km/s pc$^{2}$ for the three targets. 
We used AGNfitter (\citealt{Calistro2016}) to fit their SEDs and decompose AGN and galaxy emission, obtaining estimates of the infrared luminosity and the stellar mass of the host galaxies.

When we place these targets in the  L$'_{CO}-$L$_{IR}$ plane (Fig. \ref{LcoLir}) and compare their positions with respect to the relation expected for MS galaxies, we find that BzK4892 is the only target for which we could derive a physically stringent upper limit on the CO luminosity. 
 This source display a high star formation rate surface density, $\Sigma_{SFR}\sim 130$ M$_\odot$ yr$^{-1}$ kpc$^{-2}$ (for an effective radius $r_e=0.8$ kpc; \citealt{Barro2016}), a low gas fraction (f$_{gas}<0.06$) and a star formation efficiency SFE $>410$, pointing to an advanced level of gas depletion in this CT QSO. 
For BzK8608 and CDF153, instead, the derived upper limits on the gas fractions ($f_{gas}\lesssim 0.5$) and lower limits on the SFE of $\approx 20$, are compatible with available measurements for normal MS and SMGs.  Lower gas fractions would be inferred from our CO observations if these sources can be shown to have ISM conditions generally associated with low CO-to-gas conversion factors (e.g., high $\Sigma_{SFR}$ and/or high $T_{dust}$). We also reiterate here that, for BzK8608, all measurements should be regarded as tentative measurements because of the adopted photometric redshift (\S \ref{observations}).

In the second part of the paper, we combined our measurements with a sample comprising normal and sub-millimetre galaxies, plus dusty (see \S \ref{biasesSFE}) and unobscured AGN at z $>1$ from the literature. 
Our results suggest that sub-millimetre detected, dusty AGN are associated with star formation efficiencies very similar to those of unobscured QSOs, and higher than those of SFGs.

Furthermore, we found that CT QSOs may behave more similarly to optically luminous QSOs and obscured AGN than to SMGs: indication of outflows in the ionised and molecular phases of the ISM (SW022550, SW022513 and GMASS953) and indirect evidence of depleted gas (BzK4892) point to the possibility that, in the context of the SB-QSO paradigm, highly obscured systems are affected by feedback phenomena right from the early phases of the evolutionary sequence. 
We want to note that evidence of outflows even in early/intermediate stage merger systems has been collected for several nearby ULIRG/AGN galaxies (e.g. \citealt{Feruglio2013,Feruglio2015,Rupke2013,Rupke2015,Saito2017}) and in the 3C 298 radio-loud QSO at z = 1.439 (\citealt{Vayner2017}; see also \S \ref{mergervsorientation}; footnote \ref{SMMJ}). These results, together with the fact that SFE of dusty and unobscured QSOs are, on average, comparable, could be consistent with the SB-QSO evolutionary scenario assuming that feedback effects occur early in the evolutionary sequence. 

Our findings are not in tension with recent observational and theoretical work showing that the SFR of galaxies hosting AGN is not strongly affected by feedback (e.g. \citealt{Mullaney2015,Roos2015}): the high SFE in our dusty AGN is reasonably due to depleted cold gas reservoirs (see Fig. \ref{fgasplot}) rather than significant SFR variations. We also note that, with respect to the \citet{Mullaney2015} sample, mostly comprising MS galaxies hosting AGN, in this paper we consider starbursting, dusty AGN, which are expected to be affected by strong outflows (e.g. \citealt{Hopkins2008}).   

We also want to emphasise that particular chemical and kinematic conditions could be at the origin of the observed L$_{IR}/$L$'_{CO}$ ratios (e.g. \citealt{Blitz2006, Saintonge2016,Shi2011}), and the similarly enhanced SFE (and depleted gas reservoirs) in optically bright and dusty AGN could not necessarily imply a temporal sequence for the two samples. For example, powerful outflows, revealed both in highly obscured and optically bright QSOs (e.g. \citealt{Bischetti2017,Carniani2017,Perna2015b}), could peculiarly alter the cold gas conditions and be responsible for the enhanced SFE values. The exact interpretation of our results needs further exploration. 

We also stress that the small number of highly obscured sources and the uncertainties still affecting the measurements of both CO and IR luminosities do not allow us to draw any strong conclusion about the physical conditions in these distant buried QSOs and the possible evolutionary connection between unobscured and highly obscured AGN. Further observations of high-z SMG/QSOs with JVLA, ALMA and/or NOEMA facilities will help shed light on the exact nature of the host galaxies of rapidly growing SMBHs, both in terms of assembling larger samples, and also obtaining higher spatial resolution detections to unveil the direct effects of feedback phenomena on the host galaxy gas reservoirs.

Finally, we also want to emphasise that CO(1-0) measurements in high-z galaxies are of enormous utility to study galaxy evolution, as they are not affected by excitation correction uncertainties and can be used to compare galaxies with unknown and (possibly) different physical conditions. The CO(1-0) luminosity can be used to derive molecular gas content in the host galaxy, assuming a luminosity-to-gas-mass conversion factor. This estimate is arguably less affected by the uncertainties than measurements based on dust continuum observations, for which assumptions on the dust-to-gas ratios and/or metallicity and dust temperature are required (e.g. \citealt{Magdis2012,Scoville2016}). Existing CO(1-0) observations, however, have been so far limited to a small number of sources (see e.g. Tab. \ref{literaturesample}, for the sample of dusty AGN); the measurements reported in this paper for BzK4892 would be therefore of utility to establish the context for the next studies of highly obscured QSOs in the high-z Universe.

\vspace{2cm}

{\small 
{\it Acknowledgments:}
We thank the anonymous referee for a useful and detailed report which helped improve the paper.
 MP thanks the Marco Polo Project of the Bologna University and the University of Sussex for the hospitality during his stay in Brighton, where part of this work was carried out. MP also thanks S. Salvadori, S. Gallerani, A. Pallottini, M. C. Orofino and L. Graziani for fruitful discussions.  
GC acknowledges the support by INAF/Frontiera through the ``Progetti Premiali'' funding
scheme of the Italian Ministry of Education, University, and Research, and the INAF PRIN-SKA 2017 program 1.05.01.88.04. 
This research was finalised at the Munich Institute for Astro- and Particle Physics (MIAPP) of the DFG cluster of excellence ``Origin and Structure of the Universe'', during the program ``In \& Out: What rules the galaxy baryon cycle?'' (July 2017). MP, MB and GL acknowledge support from the FP7 Career Integration Grant ``eEASy'' (``SMBH evolution through cosmic time: from current surveys to eROSITA-Euclid AGN Synergies'', CIG 321913). MTS was supported by a Royal Society Leverhulme Trust Senior Research Fellowship (LT150041). EL is supported by a European Union COFUND/Durham Junior Research Fellowship (under EU grant agreement no. 609412). 

This work is based on observations carried out under project number W041 [PI: C. Feruglio], with the IRAM PdBI Interferometer [30m telescope], and under project number VLA/11B-060 [PI: E. Daddi], with the JVLA telescope. IRAM is supported by INSU/CNRS (France), MPG (Germany) and IGN (Spain).
The National Radio Astronomy Observatory is a facility of the National Science Foundation operated under cooperative agreement by Associated Universities, Inc. This work has made use of the Rainbow Cosmological Surveys Database, which is operated by the Universidad Complutense de Madrid (UCM), partnered with the University of California Observatories at Santa Cruz (UCO/Lick,UCSC).
}

\begin{appendix}
\section{Far-IR photometry and SED fit comparison with previous works}\label{FIRphotometry}
 
The three obscured AGN presented in this work, BzK4892, BzK8608 and CDF153, have been observed as part of the GOODS-Herschel (\citealt{Elbaz2011}) and PEP (\citealt{Magnelli2013}) surveys. However, their PACS 70, 100, and 160 $\mu$m, and SPIRE 250, 350, and 500 $\mu$m photometric data may be affected by source confusion.  
To derive their host galaxy properties, we therefore used a new  Herschel catalog presented by T. Wang et al. (in prep), which utilises an optimised algorithm (T-PHOT; \citealt{Merlin2015}) for photometry in crowded Herschel maps. T-PHOT uses minimum chi-square estimation to generate flux density estimates using galaxy positions extracted from shorter wavelength images.

In table \ref{FIRphotometrytab} we report the T-PHOT fluxes used in this work and, when available, those from the Rainbow Cosmological Surveys database\footnote{\url{https://rainbowx.fis.ucm.es/Rainbow_Database/Home.html}} for comparison.

BzK4892, BzK8608 and CDF153 multi-band SED fits are also presented in the Rainbow database. However, for CDF153 and BzK8608 the SED fits were performed excluding the un-deblended, and hence potentially biased, $\lambda > 24\mu$m photometry in the Rainbow database. Furthermore, in the Raindow SED fits only stellar SED templates are used to reproduce the photometry.
As a result, the $L_{IR}$ and SFR measurements of these two sources cannot be compared with our measurements (\S \ref{SEDsect}). We also note that the Rainbow photometric redshift of BzK8608 is z$=$ 2.63, while in this work we used z $=$ 2.51.   
The Rainbow BzK4892 SED is instead sampled also in the far-IR, and their $L_{IR}$ is compatible with our SED fit measurement within a factor of 2 (see also \citealt{Elbaz2017}). The stellar masses derived for the three targets are also compatible with ours within a factor of 2.

\begin{table}
\footnotesize
\begin{minipage}[!h]{1\linewidth}
\setlength{\tabcolsep}{4pt}
\centering
\caption{FIR photometric data}
\begin{tabular}{l|ccc}
band & BzK4892&BzK8608 & CDF153\\
\toprule
$S_{70\mu m}$ &  $2.32\pm 0.02$ & $<0.73$ & $0.65\pm 0.20$ \\
$S_{100\mu m}$&$5.64\pm0.2$ (5.67) & $< 0.72$ & $2.56\pm 0.10$ \\
$S_{160\mu m}$ & $13.71\pm 0.2$ (12.4) & $1.23\pm 0.36$ &$3.56\pm 0.19$ \\
$S_{250\mu m}$ &$25.88\pm 1.29$ (20.0) & $2.32\pm 0.62$ &-- \\
 $S_{350\mu m}$& $25.07\pm 0.62$(17.2) & $2.39\pm 0.69$ &-- \\
$S_{500\mu m}$ & $15.23\pm 1.02$ (16.3)& $<1.59$ & -- \\
\hline
log(L$_{IR}$) & $12.80\pm 0.01$ (12.43) & $11.42_{-0.70}^{+0.34}$ (12.26) & $11.70_{-0.07}^{+0.04}$ (12.00)\\
\hline
\toprule
\end{tabular}
\label{FIRphotometrytab}
\end{minipage}
{\bf Notes:} Far-IR photometric fluxes (in mJy) from T. Wang et al. (in prep.); $3\sigma$ upper limits are reported for S$_{70\mu m}$,  S$_{100\mu m}$ and  S$_{500\mu m}$ of BzK8608. CDF153 SPIRE data are not reported, as the Wang et al. procedure was not able to recover reliable measurements.   For BzK4892, we also reported in parentheses the photometric fluxes from the CANDELS Rainbow Cosmological Surveys database. No Rainbow measurements are reported for BzK8608 and CDF153. In the last row we also report the (8-1000)$\mu$m IR luminosities derived in this work and those from the Rainbow database.  
\end{table}

\section{BzK8608 photometric redshift}\label{zBzK8608}

\begin{figure}
\centering
\includegraphics[width=8cm,trim=10 130 0 110,clip]{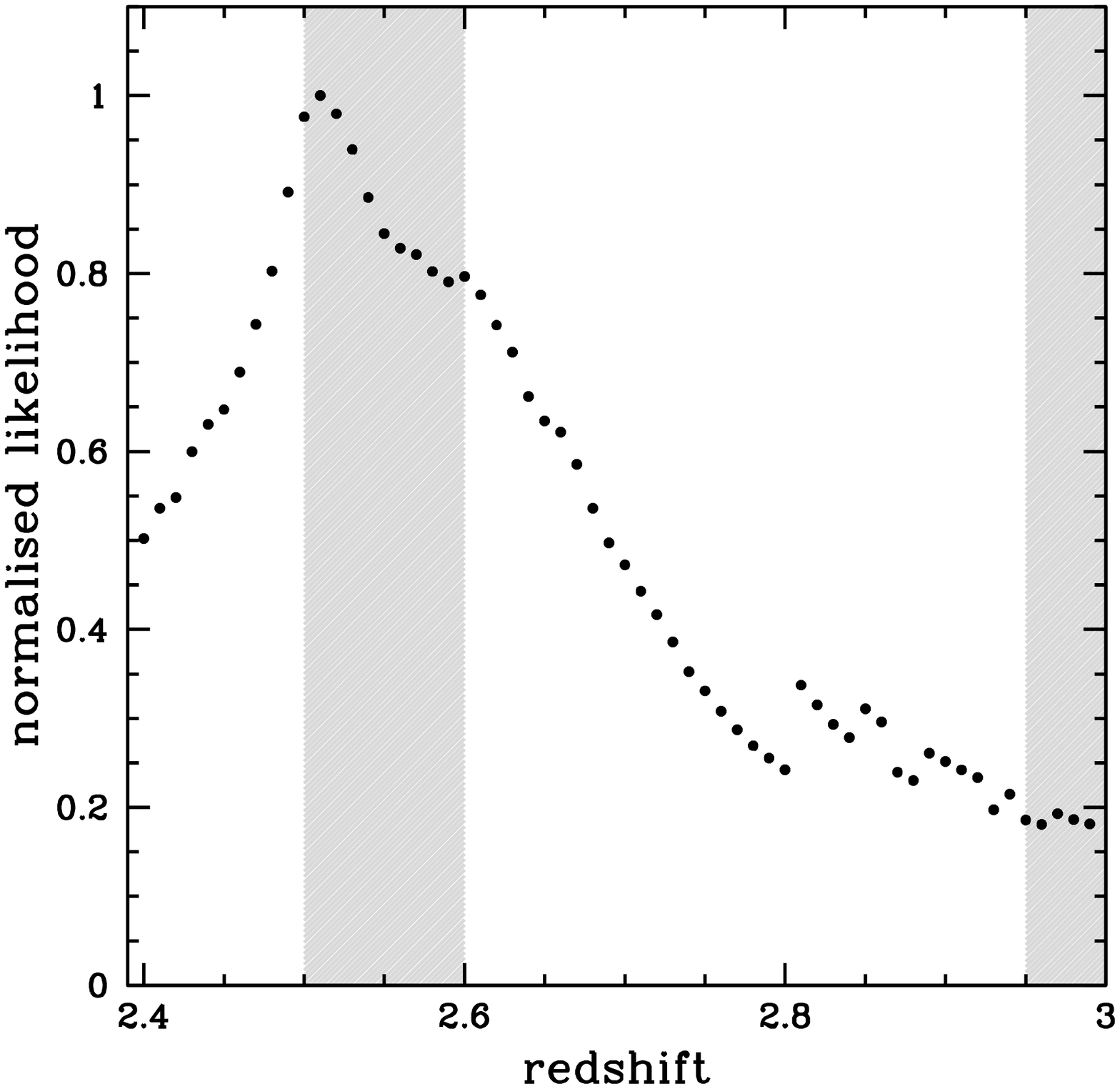}
\caption{\small  
Likelihood distribution as a function of the redshift, obtained repeating the SED fit computation of BzK8608 for fixed z value (the redshift binning is 0.01). The maximum corresponds to the redshift z =2.51. Shaded areas represent the frequency coverage of our JVLA observations.
}
\label{zphot}
\end{figure}

We derived a photometric redshift for BzK8608 by running AGNfitter on a grid of redshifts ($\Delta z$ = 0.01) spanning the redshift range 2.4$<$z$<$3, corresponding to the redshifts reported for BzK8608 in previous studies (e.g. \citealt{Feruglio2011,Hsu2014}). The normalized redshift likelihood distribution is shown in Fig. \ref{zphot}; 
 the adopted redshift z$_{phot}=2.51_{-0.06}^{+0.16}$ corresponds to the peak of the distribution (redshift errors define the likelihood ratio limits at $68\%$ confidence interval; \citealt{Dagostini2003}). This redshift estimate is also consistent with the photometric redshift reported in \citet{Hsu2014}. 

Tentative spectroscopic redshifts from iron K$\alpha$ line in X-ray spectra have been reported by \citet[][z$_{K\alpha}=2.9\pm 0.1$]{Feruglio2011} and \citet[][z$_{K\alpha}=2.83\pm 0.02$]{Buchner2014}. However, this feature is not detected in the deeper {\it Chandra} observations (see \S \ref{targets}). 

The frequency coverage of our JVLA observations maps into the two redshift intervals [2.5-2.6] and [2.95-3.08]. We searched for line emission through the accessible redshift range, finding no significant features.
Given that an unambiguous spectroscopic redshift has not yet been obtained for this target, we used the derived photometric redshift for the estimate of the CO luminosity upper limit of BzK8608 and the host galaxy properties. We note that the IR luminosity and stellar mass derived with AGNfitter do not show a significant dependence on redshift in the range 2.4$<$z$<$3.

\section{SMG and QSO samples}\label{B2}
In Table \ref{literaturesampleSMG}, \ref{literaturesample} and \ref{literaturesampleQSO} we reported the information collected from the literature for SMGs, obscured and unobscured AGN, respectively. Stellar masses, X-ray luminosities and column densities are reported for the only sources for which these quantities have been reported in literature. All luminosities are re-derived using the values tabulated in the papers mentioned in \S \ref{compilation} and considering \citet{Planck2015} Cosmological parameters; the host galaxy quantities, instead, have been corrected for the adopted Chabrier IMF. Moreover, we tabulated measurement uncertainties (at $1\sigma$ confidence level) for the only sources for which these quantities were previously reported in the literature. The stellar mass measurements of unobscured AGN are not reported in Table \ref{literaturesampleQSO} (see \S \ref{gasfraction}). Gas masses are tabulated for SMGs and dusty AGN targets, and are used to derive the gas fractions shown in Fig. \ref{fgasplot}.

The SMG/QSO sources presented in \citet{Coppin2008} and \citet{Bothwell2013} have been included in the obscured (``dusty'') AGN sample table (see \S \ref{results1}). The dusty AGN are characterised by red spectra/colors (e.g. \citealt{Banerji2017,Brusa2017,Coppin2008,Fan2017,Kakkad2017,Yan2010}) and/or high column densities (e.g. \citealt{Bothwell2013,Gilli2014,Polletta2011,Stefan2015,Vayner2017}). 

All the non-lensed luminous AGN are characterised by bright UV emission (with rest-frame absolute magnitude at 1450$\AA$ of $M_{1450}<-25$; \citealt{Banados2016,Priddey2001}), while lensed galaxies are all associated with negligible obscuration (E(B-V) $\lesssim 0.1$\footnote{\url{https://www.cfa.harvard.edu/castles/}}) and the presence of BLR emission lines in their rest-frame UV spectra (e.g. \citealt{Anguita2008,Mazzucchelli2017,Sharon2016,Shields2006,Wang2013}). 
The sample of unobscured AGN do not include the bright sources for which the IR luminosities are not available (e.g. \citealt{Carniani2017}). 

We want to note that only a few optically bright QSOs have been studied through their X-ray emission and that their characterisation requires long-exposure observations. Therefore, we cannot confidently exclude the presence of nuclear obscuration in the individual targets reported in Table \ref{literaturesampleQSO}. We note however that there are several indications pointing to the fact that the population of optically luminous QSOs are associated with X-ray spectra characterised by low column densities (N$_H\lesssim 10^{22}$ cm$^{-2}$; \citealt{Nanni2017,Shemmer2006,Vignali2005}). Therefore, we can be reasonably confident about the separation between the dusty and unobscured QSO populations presented in this work.

\begin{figure}[t]
\centering
\includegraphics[width=7cm,trim=10 130 0 110,clip]{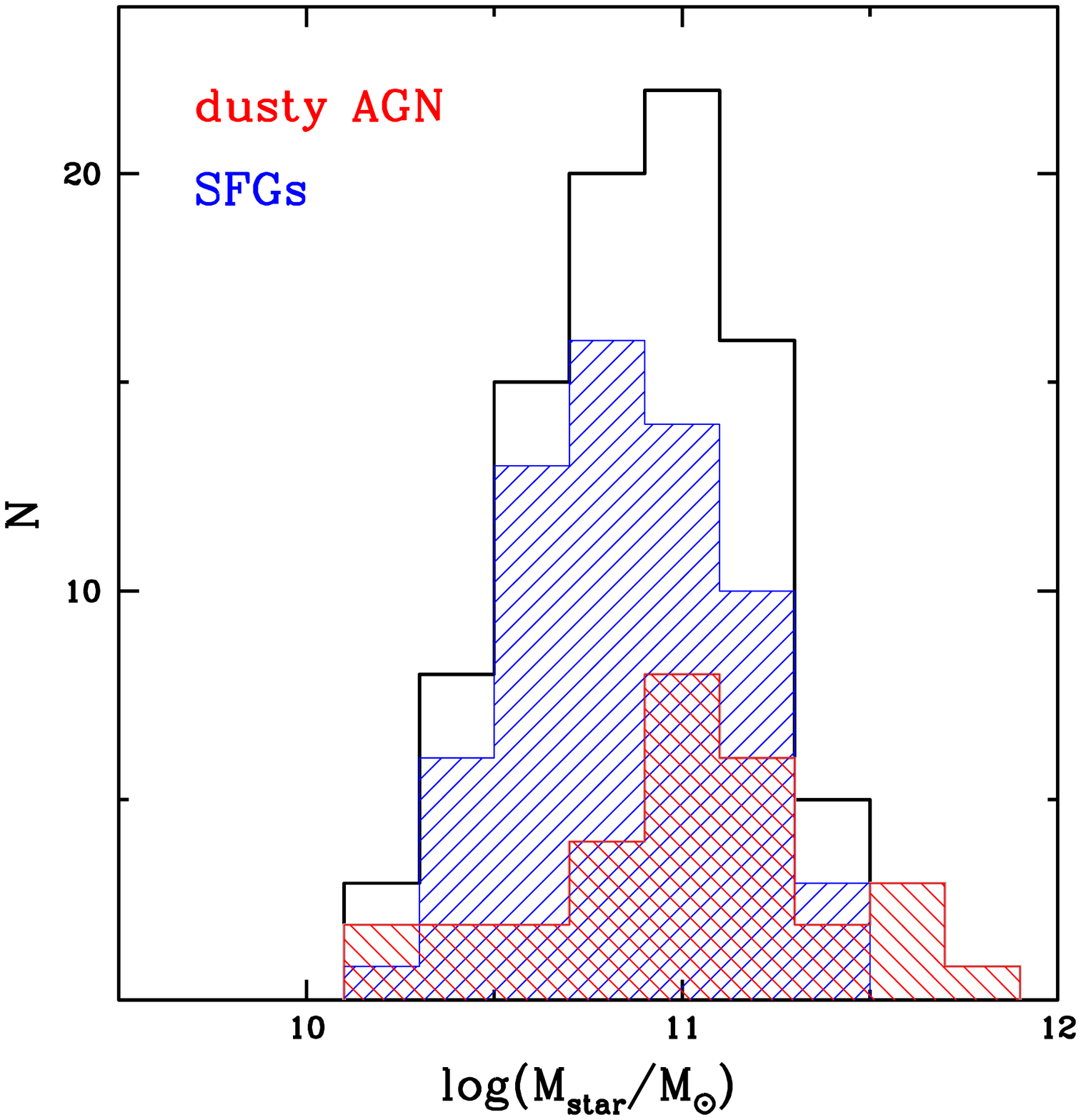}

\includegraphics[width=7cm,trim=10 130 0 110,clip]{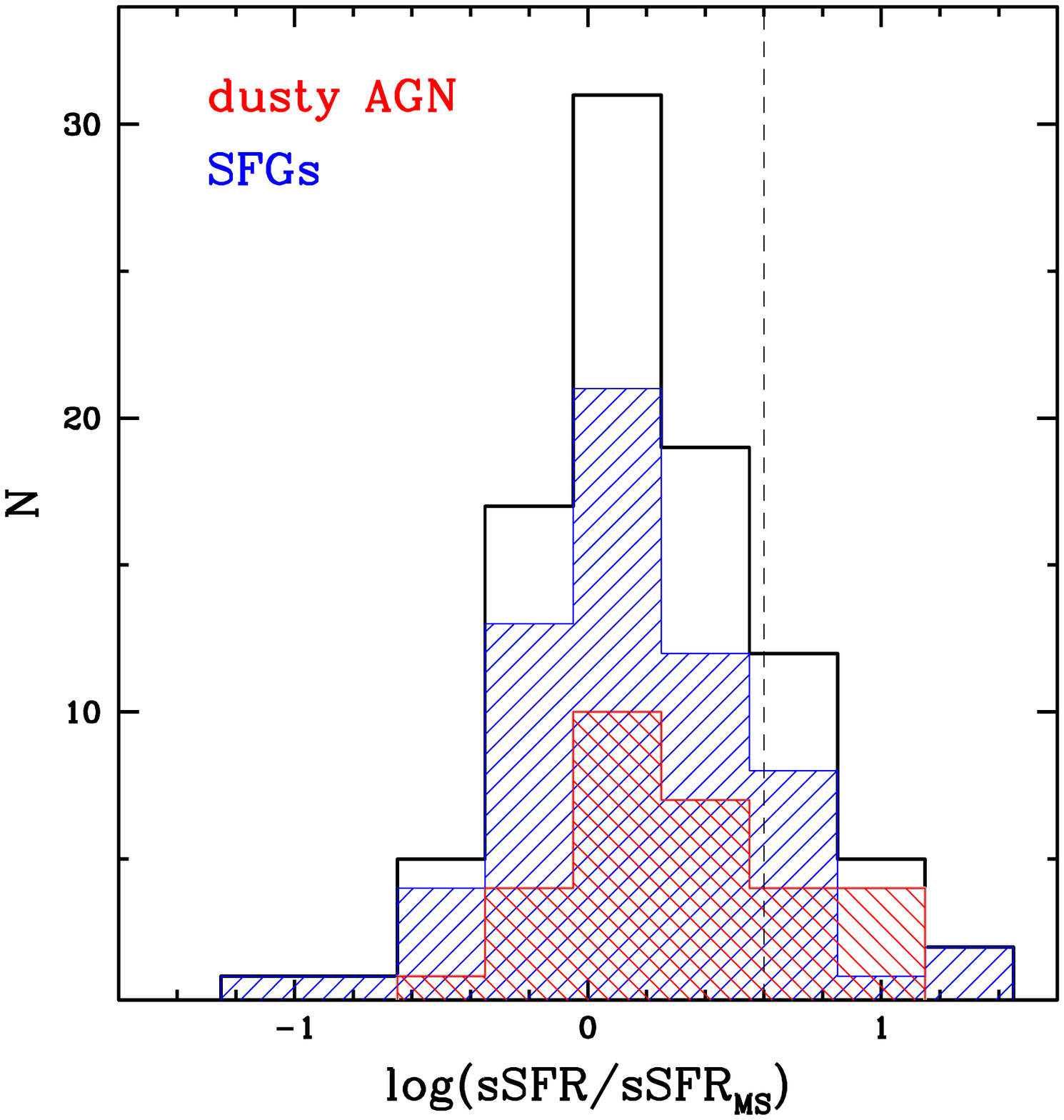}
\caption{\small  {\it Top panel}: Stellar mass distribution for sample collected in this work. Blue- and red-shaded histograms refer to the samples of SFGs (i.e. MS and SMGs) and dusty AGN. {\it Bottom panel}: specific star formation rate sSFR/sSFR$_{MS}$ distributions for the same samples in the top panel. The sSFR$_{MS}$ is derived according to the relation of \citet{Speagle2014}. In the figure, we also show the separation between normal MS galaxies and SBs at log(sSFR/sSFR$_{MS}$)$ = 0.6$.}
\label{stellarmass}
\end{figure}

 In Fig. \ref{stellarmass} we report the distribution of the stellar masses of star forming galaxies and dusty AGN; these sources are associated with massive host galaxies, with an average $\langle M_{star} \rangle =10^{11}$ M$_\odot$. In Fig. \ref{stellarmass} (bottom panel) we also report the distribution of specific star formation ratio sSFR/sSFR$_{MS}$ (according to the relation of \citealt{Speagle2014}) for the same sources, showing that the majority of SFGs and dusty AGN collected in this work are associated with MS galaxies, with the latter population showing slighly higher MS-offsets sSFR/sSFR$_{MS}$.

\begin{table*}
\scriptsize
\begin{minipage}[!h]{1\linewidth}
\centering
\caption{SMG sample}
\begin{tabular}{lcc|ccccccc}
target & RA \& DEC (J2000) & z & J$_{up}$ &r$_{J_{up},1}$& L$'_{CO}$ &M$_{gas}$& log$($M$_{star})$ & log(L$_{IR}$)  & ref\\
          &                               &    &             & & ($10^{10}$ K km/s pc$^2$)& ($10^{10}$ M$_\odot$) & (M$_\odot$)& (L$_{\odot}$) & \\
\tiny{  (1)} & \tiny{  (2) } &\tiny{  (3) } &\tiny{   (4) } &\tiny{  (5) }& \tiny{  (6) } &\tiny{  (7) } &\tiny{  (8) } &\tiny{  (9) }& \tiny{  (10) }\\

\toprule
SGP-196076           &00:03:07 -33:02:50 & 4.425& 4&0.41& $38.83\pm 3.8$ & $31.1\pm 3.0^\circ$ & & 13.44 & F17\\

SMMJ021725	                &02:17:25 $-$04:59:34&	2.292 &	4&0.41	&	4.7$\pm$0.5	&$4.7\pm 0.5^\Box$&   &  12.67	& Bo13\\ 
SMMJ021738-050339	&02:17:38 $-$05:03:39&    2.037 &	4&0.41    &	11.0$\pm$3.1  & 11.0$\pm3.1^\Box$&&	12.67	& Bo13\\%
SMMJ021738-050528	&02:17:38 $-$05:05:28&    2.541 &	4&0.41    &	9.4$\pm$2.9    &9.4$\pm2.9^\Box$ &  &  12.66        &  Bo13\\%
ADFS-27                         &04:36:57 $-$54:38:10&5.655&5&0.32&$35.9\pm 2.1$ & $28.7\pm 1.7^\circ$&& 13.46&F17\\%
SMMJ044315	                &04:43:07 +02:10:23&	2.509 &	3&0.52	&	12.1$\pm3.5$    &12.1$\pm3.5^\Box$& &	12.50       &  Bo13\\%
J044307+0210                &04:43:05 +02:09:06 & 2.509 & 1&1.00 & $1.48\pm 0.40$ &$1.18\pm 0.32^\circ$ & &$12.17\pm 0.10$ & S16\\
J04135+10277                &04:13:27 +10:27:43& 2.846 & 1&1.00 & $8.4\pm 1.7$ & $6.7\pm 1.4^\circ$ & &$13.24\pm 0.05$ & S16\\
G09-81106                     &08:49:36 +00:14:54&4.531&4&0.41& $15.3\pm 2.5$&$12.2\pm 2^\circ$ & & 13.39&F17\\%
SMMJ094303	                &09:43:03 +47:00:15&	3.346 &	4&0.41 	&	7.1$\pm$1.6  &7.1$\pm1.6^\Box$  &    &	13.25	&  Bo13\\%
PACS867                          &09:59:38 +02:28:56& 1.566 & 2&0.85 & $2.03\pm 0.16$ &$2.23\pm 0.18^\bullet$& 10.67& $12.62\pm 0.04$ & S15\\%
PACS299                          &09:59:41 +02:14:42&   1.645 &   2&0.85    &$3.21\pm 0.35$ &$3.53\pm 0.38^\bullet$& 10.44& $12.81\pm 0.03$ & S15\\ 
PACS819                          &09:59:55 +02:15:11& 1.445 & 2&0.85 & $4.15\pm 0.23$ &$2.49\pm 1.2^\dagger$& 10.61& $12.96\pm 0.01$ & S15\\ 
PACS282                          &10:00:01 +02:11:24 & 2.187& 3&0.70 & $3.20\pm 0.46$ &$3.52\pm 0.51^\bullet$& 10.88& $12. 83\pm 0.03$ & S15\\ 
PACS325                          &10:00:05 +02:19:42& 1.656 & 2&0.85 & $1.67\pm 0.26$ &$1.84\pm 0.28^\bullet$& 10.29& $12.21\pm 0.03$ &S15\\ 
PACS830                          &10:00:08 +02:19:01& 1.463 & 2&0.85 & $4.55\pm 0.35$ &$7.3\pm 2.2^\dagger$& 10.86& $12.78\pm0.01$ & S15\\ 
AzTEC-3                          &10:00:20 +02:35:22 & 5.299&2&sled& 7.6  &6.1$^\circ$& & 13.1& F17\\ 
PACS164                          &10:01:30 +01:54:12 & 1.648& 2&0.85 & $2.93\pm 0.47$ &$3.22\pm 0.50^\bullet$&$>10.28$ & $12.62\pm 0.01$ & S15\\ 
SMMJ105141	                &10:51:41 +57:19:52&	1.214 &	2&0.84	&	$6.55\pm0.52$      &$6.55\pm0.52^\Box$& &	12.69 &  Bo13\\ 
SMMJ105151	                &10:51:51 +57:26:35&	1.597 &	2&0.84	&	$2.64\pm1.45$      &$2.64\pm1.45^\Box$& 11.16& 	12.54       &  Bo13\\
SMMJ105227	                &10:52:27 +57:25:16&	2.443 &	3&0.52	 &	$2.87\pm0.61$    &$2.87\pm0.61^\Box$&   11.24&		12.82  &	  Bo13\\%
SMMJ105307  	                &10:53:07 +57:24:31&	1.524 &	2&0.84	 &	$2.21\pm0.31$  &$2.21\pm0.31^\Box$& &	  12.21  &    Bo13\\%
SMMJ123549	                &12:35:49 +62:15:37&	2.202 &	3&0.52	 &	$8.99\pm0.72$  &$8.99\pm0.71^\Box$& 11.20&		12.71  &	  Bo13\\%
SMMJ123555  	                &12:35:55 +62:09:02&	1.864 &	2&0.84	 &	$4.67\pm1.54$  &$4.67\pm1.54^\Box$& 11.11&		13.0 &	  Bo13\\%
SMMJ123634                    &12:36:34 +62:12:41&	1.225 &	2&0.84	 &	$3.95\pm0.42$  &$3.95\pm0.42^\Box$& 10.82&	 	12.61 &	  Bo13\\%
GN 20 	                        &12:37:11 +62:22:11& 4.055   & 2&1.0&$18.7\pm 5.2$&$15.0\pm 4.2^\circ$ &11.36 & 13.52 & F17\\%
SMMJ123712  	                &12:37:12 +62:13:26&	1.996 &	3&0.52	 &	$5.4\pm1.4$   &5.4$\pm1.4^\Box$ & 11.19&	12.46  &	  Bo13\\ 
SMMJ131201 	                &13:12:01 +42:42:08&	3.408 &	4&0.41	 &	$14.95\pm3.8$  &$14.95\pm3.8^\Box$ & 10.96& 	12.96  &	  Bo13\\ 
NGP-284357                   &13:32:51 +33:23:42&4.894&6&0.21& $30.2\pm 6.1$&$24.2\pm 4.9^\circ$& & 13.14 &F17\\ 
NGP-246114                   &13:41:14 +33:59:38& 3.847& 4&0.41& $14.3\pm 1.8$&$11.4\pm1.4^\circ$& & 13.21&F17\\ 
SMMJ163658+405728 	&16:36:50 +40:57:34&  1.193   & 2&0.84      & 	$2.2\pm 	0.4$ & $2.2\pm 0.4^\Box$& 10.89& 12.09 & Bo13\\ 
SMMJ163658+410523     &16:36:58 +41:05:24&	2.454 &   3&0.52     &	9.5$\pm1.0$ &9.5$\pm1.0^\Box$ & 11.01&	12.90  &	  Bo13\\ 
MIPS506                          &17:11:38 +58:38:36&  2.470 & 3&1.0& $1.61\pm 0.34$   &    $1.29\pm 0.28^\circ$    &    &  12.92 & Y10 \\
MIPS16144                      &17:24:21 +59:31:51&  2.131    & 3 & 1.0&$4.05\pm 0.95$   &   $3.24\pm 0.74^\circ$   &    & 12.75 & Y10\\
SMMJ221804 	                &22:18:04 +00:21:54&	2.517 &   3&0.52	&	9.96$\pm$1.12    &9.96$\pm1.12^\Box$&10.79 &	12.53  &	 Bo13\\%
J22174+0015                  &22:17:34 +00:15:33 & 3.099 & 1&1.00 & $4.3\pm 1.0$ &$4.3\pm 1.0^\Box$ &10.51 & $12.66\pm 0.11$ & S16\\
\toprule
\multicolumn{10}{c}{Lensed galaxies}\\
\hline
J083051.0+013224         &08:30:51 +01:32:24 & 3.634 & 3&0.17 & $10.0\pm 4.4$ &$8.0\pm3.5^\circ$& & $13.11\pm 0.12$ & Y17\\
J085358.9+015537         &08:53:58 +01:55:37 & 2.092  & 2&0.84 & $2.3\pm 0.9$&$1.8\pm 0.7^\circ$ & & $12.39\pm 0.19$ & Y17\\
J090302.9-014127         &09:03:03 $-$01:41:27 &  2.304 & 3&0.17 & $7.8\pm 2.1$&$6.2\pm 1.67^\circ$ & &$12.92\pm 0.17$ & Y17\\
J090311.6+003906         &09:03:11 +00:39:06 & 3.042 & 3&0.17 & $5.7\pm 1.5$&$46\pm1.2^\circ$ & &$12.48\pm 0.12$ & Y17\\
J114637.9-001132          &11:46:37 $-$00:11:31 & 3.259 & 4&0.41 & $8.3\pm 2.4$&$6.6\pm1.9^\circ$& & $12.97\pm 0.12$ & Y17\\
HDF 850.1                        &12:36:51 +62:12:25 &5.183&2&sled& 4.6 &$3.7^\circ$& & 12.83 & F17\\ 
J125632.7+233625          &12:56:32 +23:36:27 & 3.565 & 3&0.17 & $5.7\pm 1.5$ &$4.6\pm 1.2^\circ$ & &$12.82\pm 0.15$ & Y17\\
J133008.4+245900         &13:30:08 +24:58:59 & 3.111 & 5&0.32 & $2.4\pm 0.6$& $1.9\pm 0.5^\circ$& &$12.68\pm 0.16$ & Y17\\
J133649.9+291801         &13:36:49 +29:18:01 & 2.202 & 3&0.17 & $5.8\pm 1.6$&$4.6\pm 1.3^\circ$& & $12.89\pm 0.26$ & Y17\\
J134429.4+303036         &13:44:29 +30:30:34 & 2.301 & 5&0.32 & $5.5\pm 1.1$&$4.4\pm 0.9^\circ$& & $12.75\pm 0.12$ & Y17\\
J141351.9-000026          &14:13:51 $-$00:00:26 & 2.478 & 3&0.17 & $28.8\pm 7.6$&$23.0\pm 6.8^\circ$ && $12.95\pm 0.13$ & Y17\\
G15v2.779                       &14:24:13 +02:23:04& 4.243 & 4&0.41 & $18.0\pm 4.3$&$14.4\pm 3.4^\circ$ &&$13.10\pm 0.14$ & Y17\\
HFLS3                              &17:06:47 +58:46:23& 6.337    & 1&1.0  &      $9.49\pm 3.13$ &$9.49\pm 3.13^\Box$ & 10.57&$13.45\pm 0.05$ & F17\\  
\hline
\end{tabular}
\label{literaturesampleSMG}
\end{minipage}
Column (1): target name; Column (2): coordinates; Column (3): redshift; Column (4): $J_{up}$ refers to the upper level of the  ($J_{up} \rightarrow J_{up}-1$) CO transition; Column (5): the excitation correction defined as r$_{J_{up},1}$ = CO($J_{up} \rightarrow J_{up}-1$)/CO($1\rightarrow 0$); Column (6): CO(1-0) luminosity; Column (7): gas mass; Column (8): stellar mass; Column (9): 8-1000$\mu$m IR luminosity; Column (10): reference for the paper from which we derived CO and far-IR measurements, as well as assumed excitation correction and $\alpha_{CO}$ conversion factors. 
All lensed galaxy properties are corrected for magnification.

{\bf References:}\\
\citet[][Bo13]{Bothwell2013}; \citet[][F17]{Fudamoto2017} \citet[][S16]{Sharon2016}; \citet[][S15]{Silverman2015}; \citet[][Y10]{Yan2010}; \citet[][Y17]{Yang2017}.
\\
{\bf Notes:}\\ 
The three lensed galaxies from \citealt{Fudamoto2017} SGP-261206,  NGP- 190387 and G09-83808 are not included in our sample as their magnification factors are unknown. The S15 and Y17 infrared luminosities are derived from SFR, using the following prescription: L$_{IR} =$ SFR $\times 10^{10}$ L$_\odot$ (\citealt{Kennicutt1998}, correcting for a Chabrier IMF). \\
The luminosity-to-gas-mass conversion factors used to derive the gas mass in the different papers are indicated in the table as it follows:\\
$^\Box$: $\alpha_{CO}=1.0$;\\ 
$^\circ$: $\alpha_{CO}=0.8$;\\ 
$^\bullet$: $\alpha_{CO}=1.1$;\\
$^\dagger$: the gas masses are derived from a CO-based dynamical mass estimate, by subtracting stellar and dark matter components (\citealt{Silverman2015}). The inferred $M_{gas}$ allowed the estimate of $\alpha_{CO}=0.6\pm 0.3$ (PACS819) and $\alpha_{CO}=1.6\pm 0.6$ (PACS830).

\end{table*}

\begin{table*}
\scriptsize
\begin{minipage}[!h]{1\linewidth}
\setlength{\tabcolsep}{4pt}
\centering
\caption{obscured QSO sample}
\begin{tabular}{lcc|ccccccc|ccc}
target & RA \& DEC (J2000) & z & J$_{up}$ & r$_{J_{up},1}$& L$'_{CO}$ & M$_{gas}$& log$($M$_{star})$ & log(L$_{IR}$)  & ref& N$_H$ & log$(L_X)$&ref.\\
          &                               &    &              & &($10^{10}$ K km/s pc$^2$)& ($10^{10}$ M$_\odot$) & (M$_\odot$)& (L$_{\odot}$) & &($10^{22}$ cm$^{-2}$) & erg/s&\\
\tiny{  (1)} & \tiny{  (2) } &\tiny{  (3) } &\tiny{   (4) } &\tiny{  (5) }& \tiny{  (6) } &\tiny{  (7) } &\tiny{  (8) } &\tiny{  (9) }& \tiny{  (10) } &\tiny{  (11) } &\tiny{  (12) }  &\tiny{  (13) }\\
\toprule
ULASJ0123    &01:23:12 +15:25:22 &   2.629   &   3& 0.8&$6.8\pm 0.3$   &   $5.4\pm 0.2^\circ$   &    &   $13.24\pm 0.07$& Ba17 && &\\%
W0149+2350     &01:49:46 +23:50:14& 	3.228 &    4 &0.87 &  $2.24\pm 0.52$   &	$1.8\pm 	0.4^\circ$  & & $13.25\pm 0.05$ & F18& -- & $<45.66$ & Vi17\\
MRC0152-209    & 01:54:55 $-$20:40:26    & 1.921 &1 & 1.0 & $6.78\pm 0.82$& $5.44\pm 0.66^\circ$& 11.76& 13.26 & E14& &&\\%
W0220+0137      &02:20:52 +01:37:11&   	3.122 &	4 &0.87 &  $3.15\pm 0.66$   &	$2.52\pm 0.53^\circ$  & & $13.54\pm 0.07$ & F18& $28.2\pm 0.9$ &$44.49\pm 0.28$ & Vi17\\
SMMJ030227 &  03:02:27 +00:06:52  &1.406 & 2 & 0.84&$1.24\pm 1.44$ & $1.24\pm 1.44^\Box$ & 11.6 &12.7 & Bo13 && &\\ 
no.226	   & 03:32:16 $-$27:49:00&		1.413&	2  &0.8	&	$0.67	\pm 0.24$	&	$2.4\pm 0.9^\star$ &		10.9	&		12.22	& K17	&1.7 &41.71 &Lu17\\ 
 no.682	   & 03:32:59 $-$27:45:22&		1.155&	2	&0.8 &	$<0.84$			&	$<3.02^\star$	   &		10.6	&		11.92	& K17	 & $<1$  &42.43& Lu17\\
W0410-0913     &04:10:10 $-$09:13:05&    	3.592 &	4 &0.87 &  $16.28\pm 1.44$ &	$13.1\pm 1.15^\circ$ &  & $13.72\pm 0.03$ & F18& -- &	$<46.08$ &V17\\
X2522	   & 09:57:28 +02:25:42&		1.532&	2	&0.8 &	$<1.67$			&	$<6.00^\star$	   &		11.3	&		12.5	& K17	&$<0.8$ &45.03 &M16\\
X5308	   & 09:59:22 +01:36:18 &		1.285&	2	&0.8 &	$<1.06$			&	$<3.82^\star$	   &		11.0	&		12.02	& K17	&$<2$  & 44.28&M16\\
C1148	   & 10:00:04 +02:13:07 &		1.563&	2	&0.8 &	$<1.66$			&	$<5.98^\star$	   &		11.1	&		12.2	& K17	& $>2.7$	 & 44.42&M16\\
COSMOS22995 & 10:00:17 +02:24:52 &2.469  & 1   &  1.0&    $<1.3$                           &      $<1.3^\Box$   & 11.08     &   12.3 & Sp16 & $33\pm 17$  &43.91& Br14 \\
COSB011      & 10:00:38 +02:08:22    & 1.827   &   sled    & --&     $5.89\pm 1.16$      &     $4.71\pm 0.93^\circ$    &             &13.13&A08 & $1-10$ &44.02& A08\\
C152	   & 10:00:39 +02:37:19&		1.188&	2	&0.8 &	$<0.53$			&	$<1.91^\star$	   &		10.4	&		11.52	& K17	&$<2.6$ & 43.9&M16\\
C92             & 10:01:09 +02:22:55&		1.581&	2	&0.8 &	$2.64	\pm 0.25$	&	$9.5\pm 0.9^\star$ &		11.2	&		12.8	& K17	&$3.5_{-1.5}^{+1.8}$  &43.73& M16\\
C103	   & 10:01:10 +02:27:17&		1.433&	2	&0.8 &	$<0.66$			&	$<2.38^\star$	   &		10.7	&		11.82	& K17	&$<1.6$ & 44.15&M16\\
C1591       & 10:01:43 +02:33:31&		1.238&	2	&0.8 &	$1.67	\pm 0.4$	&	$6.0\pm 1.4^\star$ &		11.3	&		12.62	& K17	&$26.4_{-9.2}^{+15}$ &44.25& M16\\
C488	   & 10:01:47 +02:02:37&		1.171&	2	& 0.8&	$<0.42$			&	$<1.51^\star$	   &		10.3	&		11.62	& K17    & $4.3_{-3.1}^{+2.0}$ & 44.25&M16\\
XID2028       & 10:02:11 +01:37:06    & 1.592   &   sled    & --&     $1.43\pm 0.63$     &      $1.14\pm 0.50^\circ$    & $11.65\pm 0.35$    &$12.47_{-0.05}^{+0.01}$&B18  & $0.7\pm 0.02$ &45.32& P15\\
HS1002       & 10:05:17 +43:46:09&               2.102&    1     &1.0 &     $11.9\pm 3.4$                &    $9.5\pm 2.7^\circ$&                  &    $13.08\pm 0.08$&   C08,S16   &             &&\\
SDSS J1148+5251 & 11:48:16 +52:51:50&6.418 & 2 & 1.0 & $3.2\pm 0.3$& $2.6\pm 0.2^\circ$& & 13.16& St15 & $20_{-15}^{+20}$ & $45.18\pm 0.12$ & G17\\
RXJ121803  & 12:18:04 +47:08:51&               1.742&     2     & 1.0&     $<2.47$              &              $<1.97^\circ$&                  &    $12.8\pm 0.4$&   C08  & $2_{-1.0}^{+0.6}$& & S05\\
ULASJ1234    &   12:34:27 +09:07:54  & 2.503 &   3&0.8  &$4.4\pm 0.8$   &   $3.5\pm 0.6^\circ$   &    &   $13.24\pm 0.12$& Ba17 &$0.89\pm 0.09$& 45.13 &Ba14\\%
SMMJ123606 &  12:36:06 +62:10:24  &2.505 & 3 & 0.52&$2.86\pm 0.82$ &$2.86\pm 0.82^\Box$  &  10.3 & 12.81 & Bo13& $81_{-42}^{+97}$&43.52&L10\\
SMMJ123618 &  12:36:18 +62:15:51  &1.996 & 4 & 0.41&$4.6\pm 0.6$ &$4.6\pm 0.6^\Box$  &  10.7 & 12.9 & Bo13 & $<7.4$&& L10\\ 
SMMJ123707 &  12:37:07 +62:14:08  &2.487 & 3 & 0.52&$6.53\pm 1.53$ &$6.53\pm 1.53^\Box$  &  10.8 & 12.59 & Bo13 & $32.6_{13.8}^{+15.2}$ & 43.50 &L10\\  
SMMJ123711 &  12:37:11 +62:13:31  &1.995 & 3 & 0.52&$8.2\pm 1.7$ &$8.2\pm 1.7^\Box$  &  11.5 & 12.83 & Bo13 & $67.1_{-23.2}^{+41.8}$ &43.36 &L10\\
SMMJ1237   & 12:37:16 +62:03:23&               2.057&    3      &1.0 &     $<0.62$             &                $<0.5^\circ$&                  &   $12.71\pm 0.4$&    C08  &$0.17$                & 44.02  &Y09\\
RXJ1249      & 12:49:13 $-$05:59:19 &               2.240&    1     & 1.0&     $5.1\pm 1.2$                 &  $4.1\pm 1.0^\circ$    &                   &   $12.84\pm 0.08$&  C08,S16   &$0.76_{-0.32}^{+0.36} $& 44.92& S10\\
SMMJ1312   &13:12:22  +42:38:13&               2.554&    3     & 1.0&     $1.2\pm 0.4$    &     $1.0\pm 0.3^\circ$&                  &  $12.5\pm 0.4$&  C08    &     $<1.26$  &  44.6  &Bo14\\
VCV J140955      &14:09:55 +56:28:27&                2.576&    1     & 1.0&     $8.6\pm 2.3$                  &   $6.9\pm 1.8^\circ$  &                  &   $13.31\pm 0.02$&  C08,S16   &    &           &\\
3C 298        & 14:19:08 +06:28:35& 1.438 &	3	& 0.97 &       $1.15\pm 0.11$ & $0.92\pm 0.09^\circ$ &  & $12.75\pm 0.01$ & Va17&$21.4\pm 0.2$ & 46.12 & S08\\
ULASJ1539    & 15:39:10 +05:57:50   &  2.658  &    3    &1.0 &     $5.46\pm 0.94$    &     $4.36\pm 0.75^\circ$    & $10.9\pm 0.27$     &12.82& F14 & $4\pm2$ &45.12 & M17\\
J154359      & 15:43:59 +53:59:03 &              2.369&     1    &1.0 &     $2.9\pm 0.7$    &   $2.3\pm 0.6^\circ$  &                  &   $12.86\pm 0.10$&  C08,S16   & 2          &  45.46  &Y09\\
HS1611       & 16:12:39 +47:11:57 &              2.403&     3     &1.0 &     $5.2\pm 0.8$    &   $4.2\pm 0.6^\circ$  &                   &   $13.1\pm 0.4$&  C08    &     &         &\\
SMMJ163650 &  16:36:50 +40:57:34  &2.383 & 3 & 0.52&$11.4\pm 0.7$ &$11.4\pm 0.7^\Box$ & 11.0& 13.24& Bo13 &&&\\
MMJ163655   & 16:36:55 +40:59:10 &    2.607 &    3  &1.0 &     $<0.8$                 &   $<0.6^\circ$                  &                      &   $12.8\pm 0.4$&   C08& $32\pm 6$   &  44.89           &M03\\
SMMJ163706 &  16:37:06 +40:53:15  &2.377 & 3 &0.52 &$3.8\pm 0.7$ &$3.8\pm 0.7^\Box$  &  11.1 & 12.76 & Bo13 &&&\\
J164914        & 16:49:14 +53:03:16 &2.270&   3       & 1.0&     $<2.2$            &   $<1.8^\circ$                  &                    &   $12.9\pm 0.4$&  C08   & $<0.79$   &45.51  &\\
MIPS8342    & 17:14:11 +60:11:09   &1.562 & 2 &1.0& $2.32\pm 0.25$   &   $1.86\pm 0.20^\circ$    &    &  $12.56$ & Y10 & --&$<43.45$& B10\\
MIPS8196    &  17:15:10 +60:09:54  & 2.586 & 3& 1.0& $<1.70$   & $<1.36^\circ$     &    &  $13.0$ & Y10 & --&$<43.78$&  B10\\
MIPS8327    &   17:15:35 +60:28:25 &2.441   &3 & 1.0&$1.31\pm 0.21$   &  $1.04\pm 0.17^\circ$   &    & $12.84$ & Y10 & && \\
MIPS429       &  17:16:11 +59:12:13  & 2.213  &  3& 1.0& $2.97\pm 0.50$   &  $2.38\pm 0.40^\circ$       &    & $12.73$ & Y10 & &&\\
AMS12          &  17:18:22 +59:01:54 & 2.767  & 3  & 1.0& $4.4\pm 0.4$& $3.5\pm 0.3^\circ$ & $11.48\pm 0.04$& $13.52\pm 0.10$&S12 & &&\\
MIPS16080    &  17:18:44 +60:01:16  &2.007 &3 & 1.0&$2.08\pm 0.37$    &   $1.66\pm 0.30^\circ$   &    &  $12.72$ & Y10 & -- &$<43.7$ &B10\\
MIPS15949    &  17:21:09 +60:15:01  &2.122 & 3 & 1.0&$2.79\pm 0.29$   &   $2.23\pm 0.23^\circ$   &    & $12.91$ & Y10 & &&\\
MIPS16059    &  17:24:28 +60:15:33  &2.326 & 3 & 1.0&$1.80\pm 0.55$   &  $1.44\pm 0.43^\circ$    &    & $12.88$ & Y10 & &&\\
VHSJ2101    &21:01:19 $-$59:43:44 &      2.313  &   3  &0.8& $1.78\pm 0.11$   &   $1.44\pm 0.09^\circ$   &    &   $12.93\pm 0.15$& Ba17 && &\\
ULASJ2315   & 23:15:56 +01:43:50 &    2.560  &   3& 0.8& $4.31\pm 0.21$   &   $3.44\pm 0.16^\circ$   &    &   $13.61\pm 0.03$& Ba17 & $0.71\pm 0.44$&45.52& M17\\
\toprule
\multicolumn{13}{c}{Compton Thick QSOs}\\
\hline
SW022513     &  02:25:13 $-$04:39:19   & 3.427  &    4     & 1.0&    $8.4\pm 1.2$                &      $6.72\pm 0.96^\circ$  &11.3-11.6 & 12.5-13.3  & P11 & $\gtrsim 100$& 44.78& P11\\
SW022550     & 02:25:50 $-$04:21:49   & 3.867  &    4     &  1.0&    $5.8\pm 1.0$                &      $4.64\pm 0.80^\circ$  &11.2-11.9 & 12.5-13.2 & P11 & $\gtrsim 100$& $<45.5$& P11\\
XID403            & 03:32:29 $-$27:56:19    & 4.762   &   2 & 1.0 &   $2.0\pm 0.4$   &      $1.6\pm 0.3^\circ$    &  $10.8\pm 0.22$         &$12.78\pm 0.07$     & G14 & $140\pm 70$ &44.40& G14\\
GMASS953    & 03:32:31 $-$27:46:23& 2.225 & 3&0.1  &$2.1\pm 0.2$ & $1.7\pm 0.1^\circ$ & $11.3\pm 0.1$ & $12.33\pm 0.04$ & Po17 & $340_{-150}^{+370}$&43.61&DM\\ 
\hline
\end{tabular}
\label{literaturesample}
\end{minipage}

Column (1): target name; Column (2): coordinates; Column (3): redshift; Column (4): $J_{up}$ refers to the upper level of the  ($J_{up} \rightarrow J_{up}-1$) CO transition; Column (5): the excitation correction defined as r$_{J_{up},1}$ = CO($J_{up} \rightarrow J_{up}-1$)/CO($1\rightarrow 0$); Column (6): CO(1-0) luminosity; Column (7): gas mass; Column (8): stellar mass; Column (9): 8-1000$\mu$m IR luminosity; Column (10): reference for the paper from which we derived CO and far-IR measurements, as well as assumed excitation correction and $\alpha_{CO}$ conversion factors. Column (11) and (12): column density and 2-10 keV rest-frame absorption-corrected luminosities derived from X-ray spectroscopic analysis (see below). Column (13): reference for the paper from which we derived X-ray properties.

{\bf References:}\\
\citet[][A08]{Aravena2008}; \citet[][Ba17]{Banerji2017}; \citet[][Ba14]{Banerji2014}; \citet[][B10]{Bauer2010}; \citet[][Bo13]{Bothwell2013}; \citet[][Bo14]{Bongiorno2014}; \citet[][Br14]{Brightman2014}; \citet[][B18]{Brusa2017}; \citet[][C08]{Coppin2008};  \citet[][C10]{Coppin2010}; Della Mura et al. (in prep, DM); \citet[][E14]{Emonts2014}; \citet[][F18]{Fan2017}; \citet[][F14]{Feruglio2014}; \citet[][G17]{Gallerani2017}; \citet[][G14]{Gilli2014}; \citet[][K17:]{Kakkad2017}; \citet[][L10]{Laird2010}; \citet[][Lu17]{Luo2017}; \citet[][M16]{Marchesi2016}; \citet[][M17]{Martocchia2017}; \citet[][P15]{Perna2015a}; \citet[][P11]{Polletta2011}; \citet[][Po17]{Popping2017}; \citet[][S08]{Siemiginowska2008}; \citet[][S12]{Schumacher2012}; \citet[][S16]{Sharon2016}; \citet[][Sp16]{Spilker2016}; \citet[][St15]{Stefan2015}; \citet[][S05]{Stevens2005}; \citet[][S10]{Streblyanska2010}; \citet[][T18]{Talia2018};\citet[][Va17]{Vayner2017}; \citet[][Vi17]{Vito2017}; \citet[][Y10]{Yan2010}; \citet[][Y09]{Young2009}
\\

{\bf Notes:} \\
Molecular gas:\\
For each target we report the upper level of the  CO transition, $J_{up}$, and the excitation correction, r$_{J_{up},1}$ (see Notes in Table \ref{literaturesampleSMG}). For COSB011 and XID2028, the CO(1-0) luminosity has been derived extrapolating the ground-state transition flux from the QSOs SLEDs. We also indicate the luminosity-to-gas-mass conversion factors used to derive the gas mass in the different papers, as it follows:\\
$^\star$: $\alpha_{CO}=3.6$;\\ $^\Box$: $\alpha_{CO}=1.0$;\\ $^\circ$: $\alpha_{CO}=0.8$\\

The sample of dusty AGN comprises a large number of sources associated with evidence of multi-phase outflows: broad and shifted components in the [O III] emission lines (e.g. \citealt{Coppin2008,Perna2015a,Polletta2011}) as well as in molecular CO lines (e.g. \citealt{Banerji2017,Brusa2017,Fan2017,Stefan2015}) are usually found in these targets. For a small number of sources, there are also indications of ongoing mergers (W0149+2350, SMMJ030227, 3C 298). For these sources,  characterised by complex CO line profiles, we reported in the Table (and in the figures) the total molecular luminosities and masses obtained integrating over all the different kinematic components.\\

{\it Continued on the following page.}

\end{table*}

\begin{table*}
\scriptsize
\ContinuedFloat
\caption{(Notes)}
{
X-ray luminosities and Column densities:\\
X-ray luminosities refer to the 2-10 keV rest-frame absorption-corrected luminosities; column densities are derived from X-ray spectroscopic analysis for all but no.226 and no.682, for which $N_H$ is derived from the hardness ratio, following the prescriptions in \citet{Elvis2012}.
To our knowledge, the only CT sources with molecular line observations are XID403 (from C10, G14), SW022513 and SW022550 (from P11), and GMASS953 (from T18). The source C1148 (from K17), detected with 20 net counts in X-ray, is a candidate CT source (see \S \ref{results1}). The few CT sources at z $>1$ are tabulated in the lower part of the table.\\

Stellar masses and IR luminosities:\\
A08, B18, E14, F18, G14, K17, St15 (from \citealt{Leipski2013}), St17, Po17 (from T18), Va17 (from \citealt{Podigachoski2015}) and Y10 host galaxy properties are obtained with a two-component (AGN+galaxy) SED fit; their IR luminosities refers therefore to the only stellar component. \\
B13: IR luminosities are obtained from the 1.4-GHz continuum. \\
Ba17: IR luminosities are obtained using a modified greybody model assuming dust temperature and dust emissivity values ($T= [41, 47]$; $\beta=[1.6,1.95]$), and rescaling the model to match the 3mm observations.\\
C08: stellar mass from \citet{Wardlow2010}; L$_{IR}$ assuming a dust temperature of $T=40 $ K and a dust emissivity factor of $\nu^\beta$, with $\beta=1.5$, rescaling a modified greybody model to match the 850 or 1200$\mu$m photometry.\\
F14: stellar IR luminosity is obtained with a two-component (AGN+galaxy) SED fit, but is not well constrained due to the lack of data above the 22$\mu$m.\\
P11:  stellar IR luminosities are obtained with a two-component (AGN+galaxy) SED fit, but is not well constrained due to the lack of data between 24$\mu$m and 1.2mm. M$_{star}$ from H-band luminosity. \\
S12: IR luminosity is obtained by fitting the far-IR SED with a greybody model.\\
Sp16: SFR and stellar mass from 3D-HST catalog (\citealt{Momcheva2015}, i.e. without considering AGN component). The source is undetected in any Herschel/PACS or SPIRE bands. The infrared luminosity is derived from SFR, using the following prescription: L$_{IR} =$ SFR $\times 10^{10}$ L$_\odot$ (\citealt{Kennicutt1998}, correcting for a Chabrier IMF).\\ 
Y10: these sources are undetected in SPIRE bands; far-IR emission is constrained using upper limits and mid-IR {\it Spitzer} spectra. \\

}
\end{table*}

\begin{table*}
\scriptsize
\begin{minipage}[!h]{1\linewidth}
\centering
\caption{Optically bright QSO sample}
\begin{tabular}{lcc|ccccc}
target & RA \& DEC (J2000) & z &  J$_{up}$ & r$_{J_{up},1}$& L$'_{CO}$ & log(L$_{IR}$)  & ref\\
          &                               &    &             &   & ($10^{10}$ K km/s pc$^2$)& (L$_{\odot}$) & \\
\tiny{  (1)} & \tiny{  (2) } &\tiny{  (3) } &\tiny{   (4) } &\tiny{  (5) }& \tiny{  (6) } &\tiny{  (7) } &\tiny{  (8) } \\
\toprule
LBQS0018-0220 & 00:21:27 $-$02:03:33 & 2.620 & 3 & 1.0 & $5.7\pm 0.99$ & 13.54 & S05\\
J0100+2802    & 01:00:13 +28:02:25 & 6.326 & 2 & 1.0 & $1.25\pm 0.49$& $12.54\pm 0.09$ & W16\\
J0109–3047	&01:09:53 $-$30:47:26& 6.791 & 6 & 0.78	& $1.35\pm 0.22$ & $12.13_{-0.23}^{+0.07}$	& V17\\
SDSS J012958	& 01:29:58 $-$00:35:39 &	5.779 & 6 & 0.78 & $1.59\pm 0.33$ & $12.75\pm 0.07$ & CW13\\
SDSSJ020332	& 02:03:32 +00:12:29 &	5.72 & 6 & 0.8 & $<1.31$ & $12.64\pm0.11$ & W11a\\
CFHQS J021013.19$-$045620.9 & 02:10:13$-$04:56:20 & 6.438 & 2 &1.0 & $<1.28$ & $<12.28$& W11b\\
J0305–3150	&03:05:16 $-$31:50:55& 6.615& 6 & 0.78& $3.5\pm 0.3$ & $12.88_{-0.19}^{+0.01}$& V17\\
J033829 &03:38:29 +00:21:56&5.027& 5 & 1.0 & $2.7\pm 0.3$ & 13.24 & CW13\\
J083643.5+005453.3 & 08:36:43 +00:54:53& 5.774 &5 & 1.0 & $<1.9$& $<13.62$&M07\\
J0840+5624    & 08:40:35 +56:24:20    &5.844& 5 & 0.88 & $3.2\pm 0.4$ & 12.86* & CW13\\%
J0911+0027 & 09:11:48 +00:27:18& 2.372 &3 &1.0& $2.81\pm 0.52$& $12.55\pm 0.1$&CW13\\
 J0908-0034 & 09:08:47 $-$00:34:16& 2.551 &3 &1.0& $1.2\pm 0.2$&$12.5\pm 0.1$&CW13\\
J0927+2001	& 09:27:21 +20:01:23 &	5.7716 & 2 & 1.0 & $5.19\pm 0.72$ & 13.02* & CW13\\ 
MRC0943-242 &09:45:32 $-$24:28:50 & 2.923 & 1 & 1.0 & $8.45\pm 2.56$ & 12.99 & G16\\ 
BR 1033-0327  & 10:36:23 $-$03:43:19 & 4.509 & 5 & 1.0 & $<1.55$ & 13.00 & G99\\ 
J1044-0125	&10:44:33 $-$01:25:02	& 5.844 & 6 & 0.78 & $0.8\pm 0.2$ & 12.30* & CW13\\
J1048+4637    & 10:48:45 +46:37:18&6.227 & 6 & 0.78 &  $1.2\pm 0.2$ & 12.40* & CW13\\
PSS 1048+4407 & 10:48:46 +44:07:13 & 4.450 & 5& 1.0 & $<1.17$ & 13.11 & G99\\
BR 1117$-$1329 & 11:20:10 $-$13:46:26 & 3.958 & 4 & 1.0 & $<2.17$ & 13.28 & G99\\
BR 1144$-$0723 & 11:46:35 $-$07:40:05 & 4.147 & 5 & 1.0 & $<1.49$ & 13.00 & G99\\
BR 1202$-$0725 & 12:05:23 $-$07:42:33 & 4.693 & 1 & 1.0 & $10.1\pm 0.4$ & $13.35\pm 0.13$ & CW13\\
LBQS1230+1627B &12:33:10 +16:10:54 & 2.741 & 3 & 0.87 & $2.52\pm 0.82$ &   13.26&CW13\\
ULAS J131911 & 13:19:11 +09:50:51 & 6.132 & 6& 0.78& $1.9\pm 0.4$ & $13.0\pm 0.06$ & CW13\\
J1335+3533    & 13:35:50 +35:33:16   & 5.901& 6 & 0.78 & $2.2\pm 0.3$ & 12.58* & CW13\\%
BRI 1335-0417 & 13:38:03 $-$04:32:35& 4.407& 2 & 1.0 &$9.62\pm 1.57$ & 13.44 & CW13\\
J1425+3254    & 14:25:16 +32:54:10   & 5.892& 6 & 0.78 & $2.5\pm 0.5$ & 12.65* & CW13\\ 
CFHQS J142952& 14:29:52 +54:47:17& 6.183& 2 & 1.0& $4.7\pm 0.5$ & 12.83 & CW13\\
3C 318 & 15:20:05 +20:16:06 & 1.577 & 2 & 0.8 &$4.1\pm 0.8$ & 12.98& CW13\\
SDSS J162331.81+311200.5 & 16:23:31 +31:12:00 & 6.260 & 2 & 1.0 & $<2.0$ & $<13.41$ & W11b\\ 
J163033.90+401209.6 &16:30:33 +40:12:10 & 6.065 &6 & 1.0 & $<1.0$ & $<12.90$&M07\\
PG1634+706   & 16:34:51 +70:37:37&1.337 & 2 & 1.0 & $<0.99$ & 13.35 & E98\\ 
J2054-0005    &  20:54:06 $-$00:05:15   & 6.038 & 6 & 0.78 & $1.5\pm 0.3$ & 12.66* & CW13\\
J2310+1855    & 23:10:38 +18:55:20 & 6.002 & 6 & 0.78 &$4.2\pm 0.3$ & $13.25\pm 0.05$ & W13\\ 
J2348–3054	&23:48:33 $-$30:54:10& 6.902 & 6 & 0.78	& $1.47\pm 0.21$ & $12.67_{-0.22}^{+0.04}$& V17\\
\toprule
\multicolumn{8}{c}{Lensed QSOs}\\
\hline
HE 0230-2130 &  02:32:33 $-$21:17:26 & 2.166 & 3 &0.98 & $1.51\pm 0.06$ & 12.26 & R11\\
J04135+10277 &  04:13:27 +10:27:40& 2.846 & 3 & 0.98& $18.1\pm 1.1$ & 13.48 & R11\\
MG 0414+0534&  04:14:37 +05:34:42 & 2.639 & 3 & 0.98& 0.35 & 12.17 & R11\\
MG 0751+2716&   07:51:41 +27:16:32 & 3.199 & 3 & 0.98 & $1.48\pm 0.05$  &12.58 & R11\\
APM 08279+5255& 08:31:41 +52:45:17&3.912& 1 & 1.0&$2.51\pm 0.09$& 12.44 &R11\\
RX J0911+0551& 09:11:27 +05:50:54  & 2.796 & 3 & 0.98 & $0.54\pm 0.02$ & 12.46 & R11\\
BRI 0952-0115& 09:55:00 $-$01:30:07  & 4.434& 5 & 0.88 & $0.78\pm 0.01$  &12.48 &R11\\
Q 0957+561   & 10:01:20 +55:53:50  &1.414  & 2 & 1.0 & $1.99\pm 0.04$& 12.75 &R11\\
F10214+4724  &  10:24:34 +47:09:10 & 2.286 & 3 & 0.98 & $0.57\pm 0.02$ &12.64 & R11\\%
HE 1104-1805 &  11:06:33 $-$18:21:23 & 2.322 & 3 & 0.98 & $2.09\pm 0.07$  &12.27 & R11\\
BR 1202-0725&  12:05:23 $-$07:42:33  & 4.695& 1 & 1.0& $10.0\pm 0.4$  &13.35&R11\\
Cloverleaf        &   14:15:46 +11:29:44 & 2.558 & 3 & 0.98 & $4.26\pm 0.01$ &12.83 & R11\\
B1359+154     & 14:01:35 +15:13:26   & 3.239& 4 & 0.94 & $0.065\pm 0.001$  &11.07 & R11\\
B 1938+666    &   19:38:25 +66:48:53 & 2.059 & 3 & 0.98 & $0.13\pm 0.01$ & 11.28 & R11\\
PSS J2322+1944& 23:22:07 +19:44:23 &4.119& 1 & 1.0&$2.00\pm 0.04$ & 12.73 &R11\\
\hline
\end{tabular}
\label{literaturesampleQSO}
\end{minipage}

Column (1): target name; Column (2): coordinates; Column (3): redshift; Column (4): $J_{up}$ refers to the upper level of the  ($J_{up} \rightarrow J_{up}-1$) CO transition; Column (5): the excitation correction defined as r$_{J_{up},1}$ = CO($J_{up} \rightarrow J_{up}-1$)/CO($1\rightarrow 0$); Column (6): CO(1-0) luminosity; Column (7): gas mass; Column (8): 8-1000$\mu$m IR luminosity; Column (9): reference for the paper from which we derived CO and far-IR measurements, as well as assumed excitation correction factors. 
All lensed galaxy properties are corrected for magnification.

{\bf References:}\\
\citet[][CW13]{Carilli2013}; \citet[][E98]{Evans1998}; \citet[][G99]{Guilloteau1999}; \citet[][G16]{Gullberg2016}; \citet[][R11]{Riechers2011}; \citet[][S05]{Solomon2005}; \citet[][V17]{Venemans2017}; \citet[][W11a]{Wang2011a}; \citet[][W11b]{Wang2011}; \citet[][W13]{Wang2013}; \citet[][W16]{Wang2016}.\\
{\bf Note:}\\
Most of the luminosity measurements are from \citet[][see their Supplemental Table for details]{Carilli2013} and \citet{Riechers2011}; the luminosities marked with a * are from \citet{Wang2010} and are corrected for AGN far-IR emission. V17, W11a, W13 and W16 infrared luminosities are derived from modified greybody models, while G16 L$_{IR}$ is obtained from  a two-component (AGN+galaxy) SED fit. The IR luminosities of G99 targets are taken from \citet[][for the BR targets]{Priddey2001} and \citet[][for PSS 1048+4407]{Isaak2002} and are obtained from modified greybody models; the CO(1-0) luminosity upper limits are here inferred by assuming r$_{J_{up},1}$ = 1. W11b and E98 IR luminosities are taken from \citet{Lyu2016}, and are obtained  from  a two-component (AGN+galaxy) SED fit. LBQS0018-0220 CO and IR luminosities are reported in \citet{Solomon2005}, but no further information is found in the literature for this target. 
\end{table*}
\end{appendix}

\end{document}